\def\AuB{{Au$_{\rm B}$\ }\/}
\def\Auone{{Au$_{\rm A1}$\ }\/}
\def\Autwo{{Au$_{\rm A2}$\ }\/}
\def\Snone{{Sn$_{\rm A1}$\ }\/}
\def\Sntwo{{Sn$_{\rm A2}$\ }\/}

\documentclass[prb,preprint,floatfix]{revtex4}
\pdfoutput=1 
\usepackage{epsf}
\usepackage{psfig}
\usepackage{graphicx}

\begin{document}
\title{Electronic structure of Au-Sn compounds grown on Au(111)} 
\author{Pampa Sadhukhan$^1$, Sajal Barman$^1$, Tufan Roy$^{2,3,*}$, Vipin  Kumar Singh$^1$,  Shuvam Sarkar$^1$,  Aparna Chakrabarti$^{2,3}$ and Sudipta Roy Barman$^{1}$}
\affiliation{$^1$UGC-DAE Consortium for Scientific Research, Khandwa Road, Indore, 452001, India}
\affiliation{$^2$Homi Bhabha National Institute, Training School Complex, Anushakti Nagar, Mumbai  400094, Maharashtra, India}
\affiliation{$^3$Theory and Simulations Laboratory,  Raja Ramanna Centre for Advanced Technology, Indore 452013, Madhya Pradesh, India}

\begin{abstract}
		  The electronic structure of  Au-Sn  intermetallic layers of different compositions grown on Au(111)  to the thickness of several nanometers has been studied in this work. The layer, interface and the substrate related components in the Au 4$f$ and Sn 4$d$ core-level spectra obtained using x-ray photoelectron spectroscopy (XPS) vary with deposition parameters to reveal the details of the Au-Sn  formation. While AuSn is grown by deposition at room temperature, Au rich compounds form as a result of  heat treatment through inter diffusion of Au and Sn. Deposition at high temperature forms more Au rich compositions compared to post annealing at the same temperature due to the kinetic energy of the impinging Sn atoms in the former case. Post annealing, on the other hand, stabilizes the bulk phases such as AuSn and Au$_5$Sn and exhibits an activated behavior for transition from the former to the latter with increasing temperature. 
		  ~The XPS valence band spectra of  AuSn and Au$_{5}$Sn layers show good
		  ~agreement with the density functional theory  calculation, indicating that these have the bulk structure reported in literature. However, the influence of anti-site defects is observed in Au$_5$Sn. Low energy electron diffraction study reveals that although the AuSn layer is ordered,  its top surface is disordered at room temperature. Surface order is obtained  by annealing or deposition at elevated temperatures and   dispersing bands  are observed  by angle resolved photoemission spectroscopy. Both electron-like and hole-like bands are evident for the ($\sqrt{3}$$\times$$\sqrt{3}$)R30$^{\circ}$ phase, while a nearly free electron-like parabolic surface state is observed for the p(3$\times$3)R15$^{\circ}$ phase. 
\end{abstract}
\maketitle 

\section{Introduction}
             Interest in the electronic structure of Sn adlayers has increased many fold after the discovery of stanene, a two dimensional free standing Sn layer that behaves as quantum spin Hall insulator with a gap of 0.3 eV.\cite{Xu13} 
             Subsequently, stanene has been grown on substrates such as  Bi$_2$Te$_3$(111)\cite{Zhu15} and InSb(111).\cite{Xu17}
             ~However, while intermixing or alloying might be considered detrimental for stanene formation, a density functional theory (DFT) study by Chuang {\it et al.} predicted that surface alloys can be used to obtain 2D topological insulators.\cite{Chuang16}  
             In addition, surface alloying is important in  designing catalysts by improving their surface activity.\cite{Chen05}

          Surface alloying commonly occurs in bimetallic systems by metal deposition on a substrate with intermixing of the adatoms and the substrate atoms. For example, in systems such as Sn/Pt, Sn/Ni, Sn/Cu, Mn/Al, Al/Mn, Ge/Au, Pd/Au  surface alloy formation has been reported.\cite{Overbury92,Biswas07,Dhaka11,Wang17,Koel92}
         Surface alloy formation in Sn/Au  is particularly favored because the high diffusivity of Au  even at room temperature.\cite{Tu74,Buene78,Nakaiiara81,Barthes81,Zhang91}
          ~Based on low energy electron diffraction (LEED), Auger electron spectroscopy (AES) and electron energy loss spectroscopy it was shown  
           that deposition of Sn  on Au(111)  at room as well as at low temperature (173 K) resulted in  surface alloy formation.\cite{Barthes81,Zhang91,Kasukabea07} 
                     ~Interdiffusion in Au/Sn thin films has been investigated by other groups, who reported formation of Au$_5$Sn, AuSn, AuSn$_{2}$, AuSn$_4$ compounds at the interface depending on the growth conditions and film thickness.\cite{Dyson66,Buene78,Nakaiiara81,Hugsted82,Tang08,Paul04,Ghosh10,Yang10}  
               ~In fact, intermixing at the surface  is also expected from  the Au-Sn binary phase diagram that shows high miscibility in the bulk and 
              different intermetallic compounds have been reported.\cite{Ciulik93} 
~             ~The Au-Sn alloys have been reported in literature to be non-toxic soldering materials with good bonding characteristics and these are widely used in advanced electronic circuits as well as in optoelectronic devices.\cite{Tang08,Wronkowska13,Rerek18} 
        ~     Theoretical studies\cite{Mier17,Canzian10} also support surface alloying in Au-Sn, for example,   Meier and Castellani\cite{Mier17}  show that  at 0.5 ML coverage,  half of the Au atoms of the top layer are displaced forming a plane in between the adsorbed Sn atoms and the Au top layer atoms. This result supports an earlier Monte Carlo based theoretical study that showed a close competition between  surface alloying and layer-by-layer growth at room temperature.\cite{Canzian10}
             
  			In this paper, we present a study of the electronic structure of  Au-Sn  intermetallic compounds grown on Au(111) using x-ray photoelectron spectroscopy (XPS), LEED, and angle resolved photoemission spectroscopy (ARPES).      
             An analysis of the XPS core-level spectra shows that AuSn grows on Au(111)  at room temperature, while high temperature treatment  forms Au rich phases, for example, Au$_5$Sn. The two methods of high temperature treatment, namely high temperature deposition and post annealing have been compared.
             Our DFT calculations are in good agreement with the XPS valence bands of  both AuSn and Au$_5$Sn, and  the effect of anti-site defects is visible in the latter. 
                   ~The surface ordering of the intermixed layer is achieved by high temperature treatment that leads to  sharp LEED patterns and dispersing bands in ARPES.  

\section{Methods}
     
         Polished Au(111) surface was cleaned by repeated cycles of sputtering with 1.5 keV Ar$^{+}$ ion for 15 min and annealing at 673 K for 10 min {\it in-situ}.  
        ~ Sn was deposited on Au(111)
        ~using water cooled Knudsen cell\cite{Shukla04} operated at  1140$\pm$5 K at the chamber base pressure of 8$\times$10$^{-10}$mbar. Note that the deposition rate used here is 8 times larger than that used in our earlier work where small amount of Sn equivalent to about a monolayer was deposited using a cell temperature of 1080~K.\cite{Sadhukhan_archive} 
        
         
        Low energy electron diffraction (LEED) patterns were recorded using a  four-grid rear view optics from OCI Vacuum Microengineering. X-ray photoemission spectroscopy (XPS) measurements were carried out  using R4000 electron energy analyzer and monochromatized AlK$\alpha$ laboratory x-ray source with 1486.6 eV photon energy (h$\nu$) from  Scienta Omicron GmbH at the base pressure of 1$\times$10$^{-10}$ mbar.  The spectra were recorded at normal emission geometry with 100 eV pass energy, transmission lens mode and 0.3 mm slit width. The instrumental resolution was 0.34 eV, which was obtained  by fitting the Au Fermi edge with a Fermi function convoluted with a Gaussian function. Angle resolved photoelectron spectroscopy (ARPES) was carried out by using monochromatized He\,I source with  h$\nu$= 21.2 eV (unless otherwise mentioned) with an overall energy resolution of 100 meV. The angular resolution of 1$^{\circ}$, a   pass energy of 10 eV and a wide angle lens with an acceptance angle of $\pm$15$^{\circ}$ were used.
         All the measurements were carried out at room temperature (RT)  that varied from 300 to 320 K. The high temperature treatment  was performed by deposition at higher substrate temperatures ($T_S$) or by post annealing for  10 min to higher annealing temperatures ($T_A$) after deposition at RT.

 The thickness ($d$) of the AuSn layer has been estimated by two different methods. Here, we discuss the method based on  the  well known ($\sqrt{3}$$\times$$\sqrt{3}$)R30$^{\circ}$ phase, which  is a typical surface alloy phase with a composition of S$_2$A for most binary metallic systems, where S is substrate atom and A is the adatom.\cite{Yuhara18, Maniraj18,Osicki13}   It   corresponds to 1/3 ML of Sn that forms for a deposition time ($t_d$) of 0.5 min at RT. Therefore,  Sn deposition for $t_d$= 0.75 min corresponds to an equivalent of 0.5 ML Sn. But,  due to intermixing, formation of AuSn occurs at RT, as is evident from the XPS study discussed in Section IIIA.  1 ML of AuSn  forms  for this deposition having 0.5 ML of Sn and 0.5 ML of Au. Thus, $t_d$= 0.75 min deposition will correspond to 1 AuSn monolayer that has a thickness of 0.28 nm, as obtained from scanning tunneling microscopy (STM) image [Fig.~S1 of  the Supplementary material (SM)].\cite{supplement} Thus, if we assume that the sticking coefficient  remains unchanged, $t_d$= 57 min Sn deposition would have $d$=  (57$\times$0.28)/0.75 nm= 21.3 nm. However, as per previous Auger electron spectroscopy study,\cite{Zhang91} the sticking coefficient becomes half after the completion of the first AuSn layer. Therefore, $d$ is estimated to be $\approx$10.7~nm for $t_d$= 57 min.  We have also used a different approach to determine $d$, based on XPS, to cross-check the above discussed LEED based method, and this is discussed later in Section IIIA. 

The least square curve fitting of the core-level spectra were performed by using Doniach-$\breve{S}$unji$\acute{c}$ (DS) lineshape\cite{Doniach70} for each component with fixed Lorentzian lifetime broadening ($\gamma$) equal to that of the bulk metal. Next, each of the components were convoluted with a Gaussian function of fixed width (0.34~eV) to represent the instrumental broadening as in our earlier work.\cite{Sadhukhan19} In case of Au 4$f$, position of the bulk component was kept fixed to that of bulk Au metal. The  positions of the other components were varied allowing an extra Guassian broadening  to account for disorder related broadening. Moreover, we have also included a Tougaard  background in the fitting scheme.\cite{Tougaard89} 
~ The  error in the determination of the composition by XPS has contribution from the experimental uncertainty as well as the multi parameter  non-linear fitting.  
~Although we vary all the parameters (16 (11) parameters for the Au 4f (Sn 4d) spectrum), we impose certain constraints ($e.g.$ the Doniach-Sunjic asymmetry parameter($\alpha$)\cite{Doniach70} was taken to be the same for both the spin-orbit components and the spin-orbit splitting  was allowed to vary within a constraint of maximum 1\%) to arrive at the final converged parameters.  We have estimated the error by performing the fitting with different constraints and methods, as well as more than one experiments at same temperature and we find it to be about 15\%. 

       First principles calculations based on DFT have been performed using Vienna Ab initio Simulation Package\cite{VASP} in combination with the projector augmented wave (PAW) method\cite{PAW} including the spin-orbit coupling (soc). For the exchange correlation functional, we have used the generalized gradient approximation (GGA).\cite{PBE} We have used an energy cutoff of 500\,eV for the plane waves. The final energies have been calculated with a $k$-mesh of 10$\times$10$\times$10 for  AuSn, which is reported to possess hexagonal crystal structure and for Au$_{5}$Sn, which has trigonal structure a $k$-mesh of dimension 10$\times$10$\times$4 has been used. The energy convergence criterion has been set to 
       ~10 $\mu$eV. 
       
      The DFT calculations were performed for AuSn and Au$_{5}$Sn using their bulk crystal structure reported experimentally.\cite{Jan63,Osada74}  AuSn crystallizes in NiAs-type hexagonal crystal structure with 4 atoms in the unit cell, Au and Sn atoms being at Wykoff positions of 2a (0, 0, 0) and 2c (0.3333, 0.6667, 0.25), respectively [Fig.~\ref{AuSn_DOS}(a)]. It has P6$_{3}$/mmc space group (number 194) 
      ~with  lattice constants $a$= $b$= 4.322\AA, and $c$= 5.523\AA. Au$_{5}$Sn is reported to possess trigonal crystal symmetry with R3 space group (no. 146) with the lattice constants $a$= $b$= 5.092\AA, and $c$= 14.333\AA. The unit cell of Au$_{5}$Sn has 18 atoms in the unit cell with Au atoms at Wykoff positions of 3a (0, 0, 0.6693), 3a (0, 0, 0.3307) and 9b (0.3333, 0.3403, 0.1666), respectively; whereas Sn atoms are placed at the 3a (0, 0, 0)  [Fig.~\ref{Au5Sn_DOS}(a)]. 
      ~The  anti-site defect in AuSn has been introduced by site-exchange of one Au atom at (0, 0, 0.5) with one Sn atom at (0.6667, 0.3333, 0.75) per unit cell, while for Au$_5$Sn two types of site exchanges have been considered:  one between one Sn atom in 3a and one Au atom in 3a sites (3a-se-Au$_5$Sn) and another between one Sn atom in 3a and one Au atom in 9b sites (9b-se-Au$_5$Sn).
             
             In order to calculate the theoretical XPS valence band, all the  $s$, $p$ and $d$ partial density of states (PDOS) of Au and Sn $e.g.$  Au 5$d$, Sn $5s$, and Sn $5p$  
				~were multiplied by the corresponding photoemission cross sections of 0.0026, 0.0006 and 0.000385 mega barn, respectively for 1486.7 eV photon energy.\cite{Yeh85}
				~These were then added and multiplied by the Fermi function at 300 K and subsequently convoluted with a Gaussian function representing the instrumental resolution and an energy dependent Lorentzian function.\cite{Barman95} Furthermore, a Shirley background\cite{Shirley72} has been added to the total calculated VB.

     \begin{figure*}[htb]
     	\centering
     	\vskip -5mm
     	\includegraphics[width=170mm]{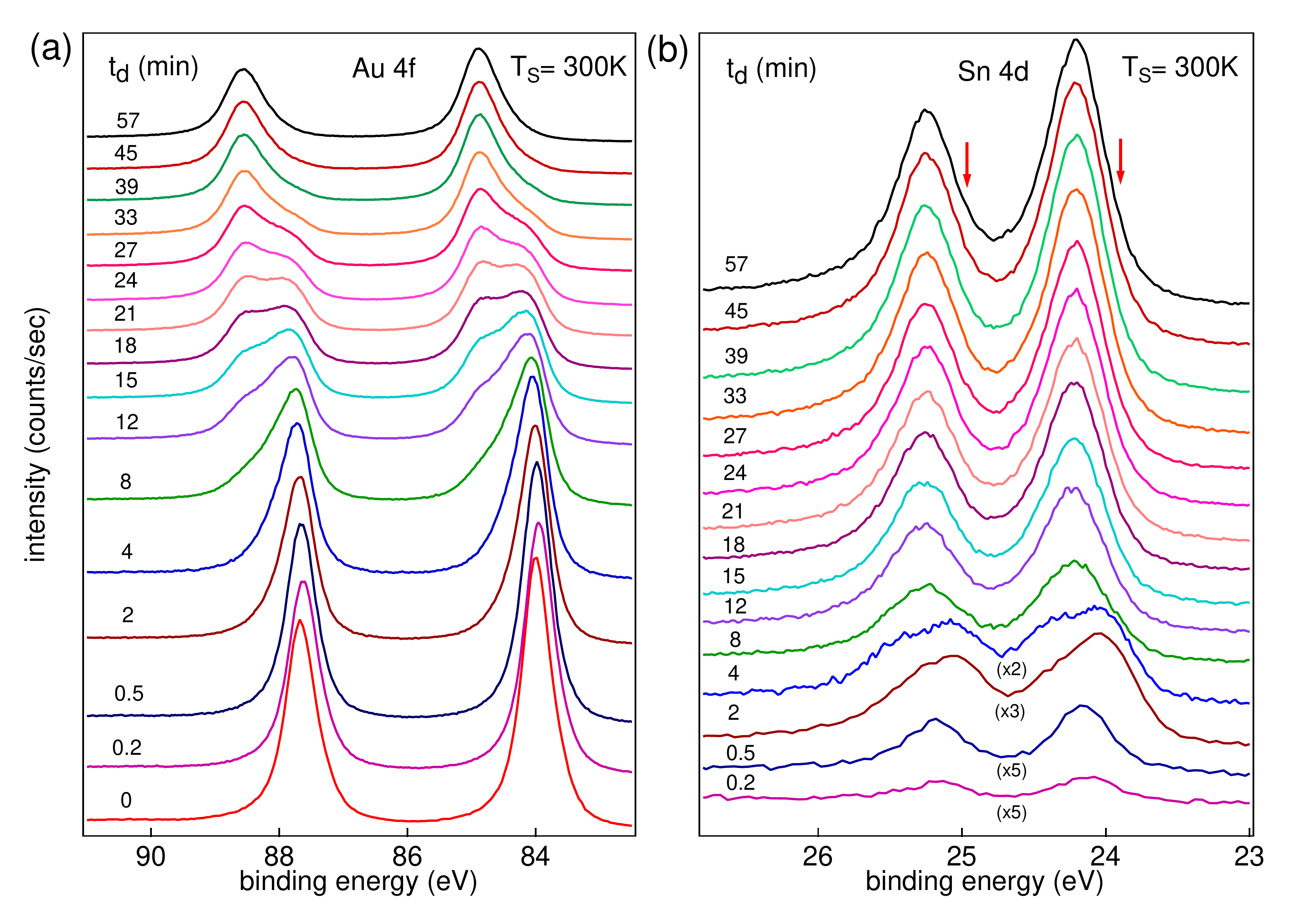} 
     	\caption{(a) Au 4$f$ and (b) Sn 4$d$ core-level spectra as a function of deposition time ($t_{d}$) of Sn on Au(111) at RT using monochromatic AlK$_{\alpha}$ radiation. The spectra have been staggered along the vertical axis for clarity of presentation. The spectrum of Au 4$f$ for $t_d$= 0 corresponds to that of bulk Au metal. In (b), the red arrows indicate the position of Sn 4$d$ peak for bulk Sn metal.} 
     	\label{core_cov}
     \end{figure*}      
     
     \begin{figure*}[htb]
     	\centering
     	\vskip -5mm
     	\includegraphics[width=170mm]{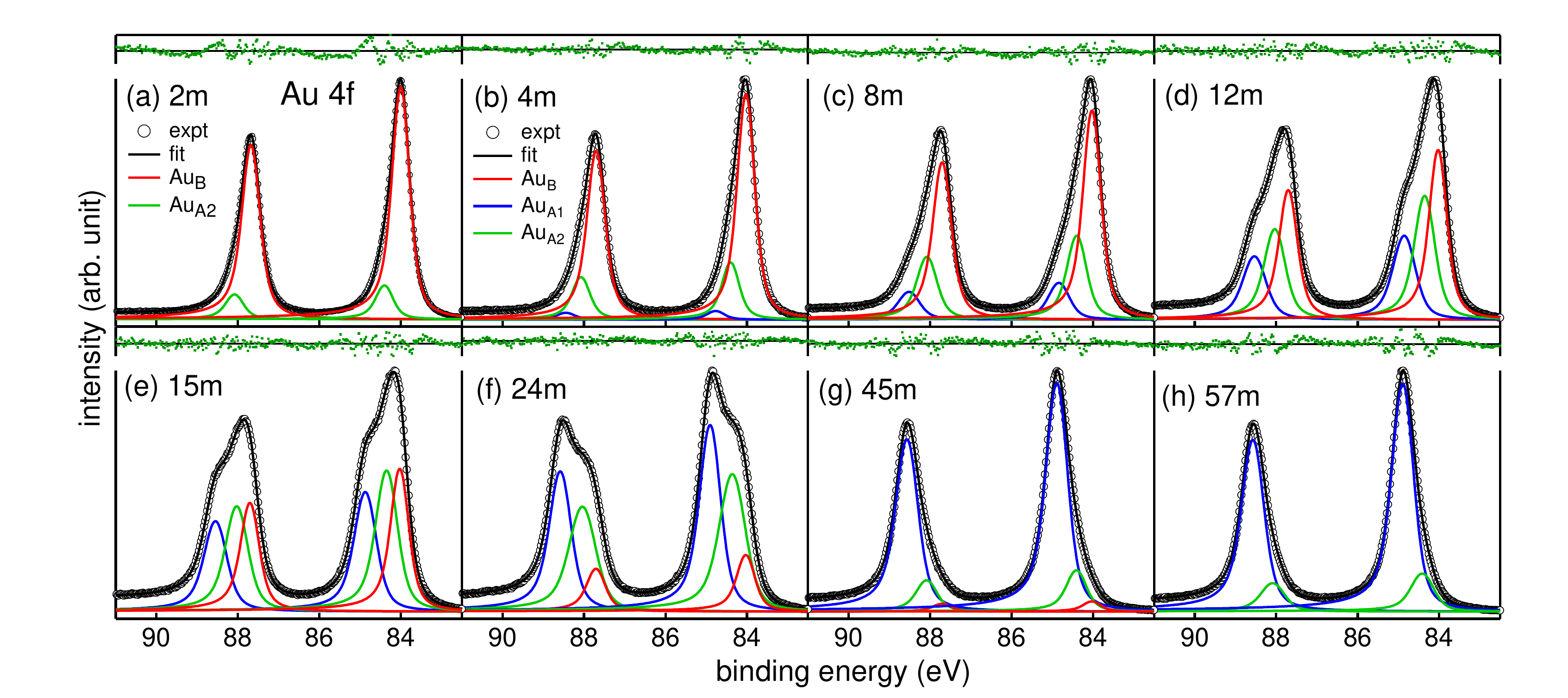} 
     	\caption{Au 4$f$ core level spectra (black open circles) that have been fitted (black line) using a least-squares  error
     		minimization routine. The spectra have been normalized to the same height. The different components obtained from the fitting are Au$_{\rm B}$ (red line), Au$_{\rm A2}$ (green line) and Au$_{\rm A1}$ (blue line) for Sn deposition on Au(111) at RT for $t_d$= (a) 2 min, (b) 4 min, (c) 8 min, (d) 12 min, (e) 15 min,  (f) 24 min, (g) 45 min and (h) 57 min.} 
     	\label{Au4fcorefit}
     \end{figure*}

          \begin{figure*}[htb]
          	\centering
          	\vskip -5 mm
          	\includegraphics[width=170mm]{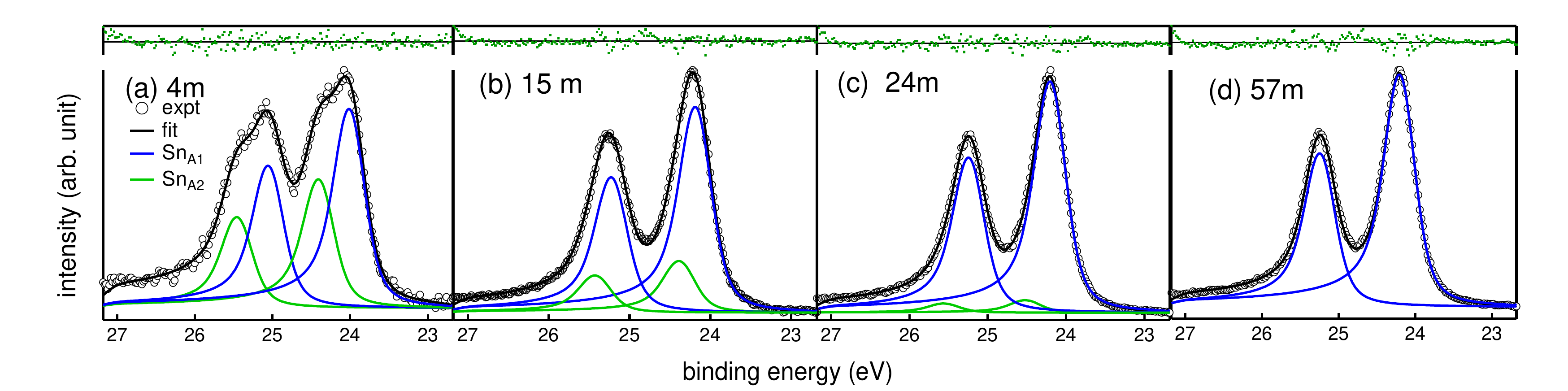} 
          	\caption{Sn 4$d$ core level spectra (black open circles) that are normalized to same height  have been fitted (black line) using a least-squares  error
          		minimization routine along with the different components, Sn$_{\rm A1}$ (blue line) 
          		and Sn$_{\rm A2}$ (green line) for $t_d$= (a) 4 min, (b) 15 min, (c) 24 min and (d) 57 min Sn deposition at RT.} 
          	\label{Sn4dcorefit}
          \end{figure*} 
          
           \begin{figure}[htb]
           	\centering
           	\includegraphics[width=170mm]{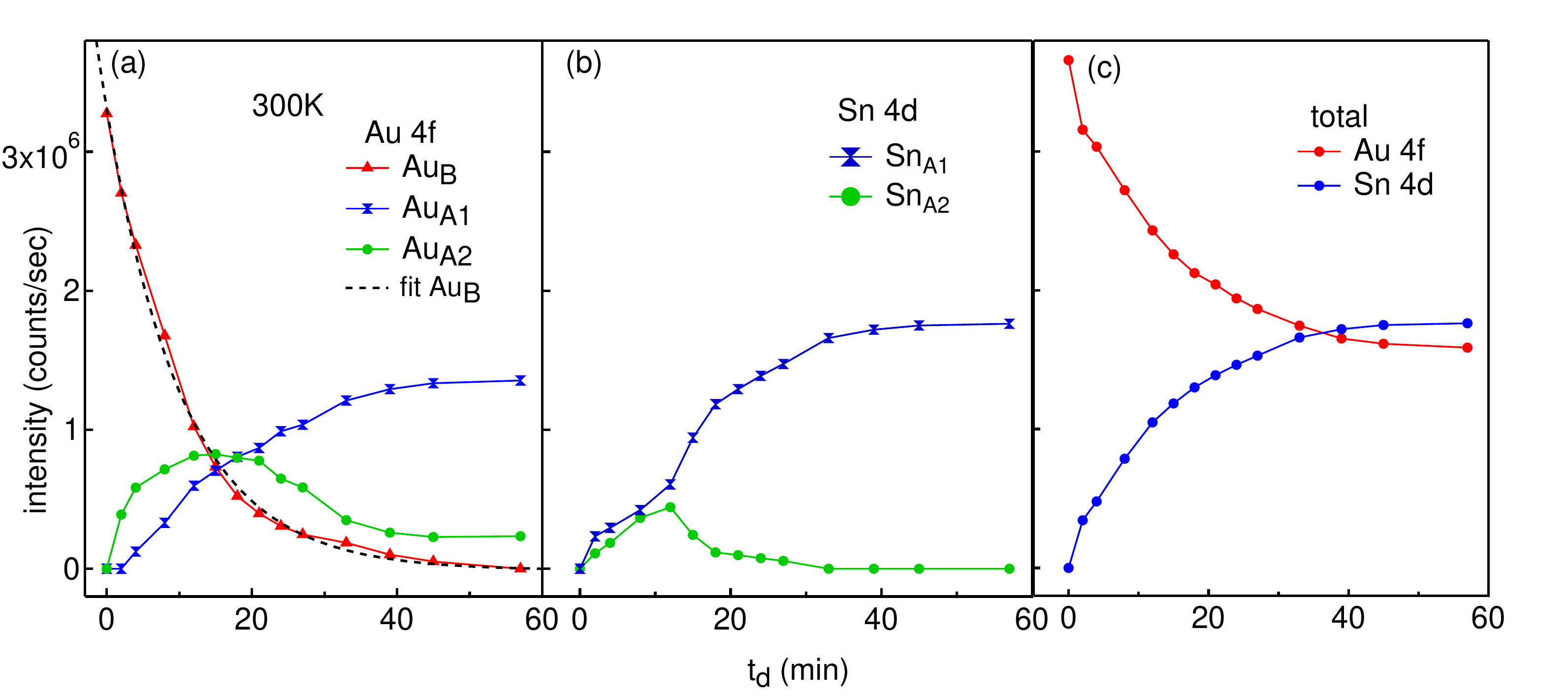} 
           	\vskip -5mm
           	\caption{Intensity as a function of $t_d$ after dividing by the respective photoemission cross-sections for the different components of (a) Au 4$f$, (b) Sn 4$d$ obtained from the least square fitting and (c) total Au 4$f$ and Sn 4$d$ intensities. An exponential fit to Au$_{\rm B}$) is shown by a black dashed line in (a).} 
           	\label{intensities}
           \end{figure}

        \section{Results and discussions}

        \subsection{XPS core-level spectra as a function of Sn deposition on Au(111) at RT}
         
         Au 4$f$ and Sn 4$d$ core-level spectra as a function of the deposition time ($t_{d}$) of Sn is shown in Fig.~\ref{core_cov}. For  $t_{d}$ ($\leq$0.5~min), the main peak of Au 4$f$ does not exhibit any discernible  change in shape and position [Fig.~\ref{core_cov}(a)]. 
          ~With increasing Sn exposure ($t_d$$\geq$2~min), the Au 4$f$ peak  shows an asymmetry on the higher binding energy (BE) side in both the spin-orbit components, which gradually develops  into  a separate peak that is about 0.9 eV shifted. For larger $t_{d}$, the overall intensity of the Au 4$f$ main peak decreases, however the peak at higher BE is relatively more intense as the main peak decreases in intensity.  In contrast, Sn 4$d$ exhibits a different behavior. The centroid of Sn 4$d$ peak is shifted by 0.25 eV from that of metallic Sn [indicated by  red arrow in Fig.~\ref{core_cov}(b)], which implies intermixing throughout the deposition range. Similar lineshape for Au 4$f$ and Sn 4$d$ have also been observed for high temperature deposition at $T_S$= 373~K, although the thickness was underestimated in that case.\cite{DAEpaper} 
         
         In metallic Sn, the plasmon related loss features are prominent in XPS. For Sn 4$d_{5/2}$, the loss features related to first bulk plasmon and the surface plasmon  appear at 38.2~eV and 34.3~eV, respectively. In contrast, the plasmon loss features are clearly suppressed  over the whole Sn deposition range  of $t_d$= 4-57 min at RT [see Fig.~S2(a)].\cite{supplement}  This implies intermixing of Sn and Au, since  if Sn grew as layer by layer, the plasmons would have appeared when a monolayer is formed, as discussed in our earlier work on thin ($<$ 1 nm) layers of Au-Sn.\cite{Sadhukhan_archive}

\underline{{\bf Least square fitting of the core-level spectra:}}         In order to extract quantitative information from the core-level spectra in Fig.~\ref{core_cov}, we have performed least square curve fitting. In Fig.~\ref{Au4fcorefit}(a) for $t_d$= 2 min, we find that the main Au 4$f_{7/2}$ and 4$f_{5/2}$ peaks appear at 84~eV and 87.7 eV, respectively that is similar to that of bulk Au metal, as also shown in Fig.~\ref{core_cov}(a). Thus,  this peak represents the signal from the Au substrate.  However, an asymmetry is observed  on the higher BE side of both Au 4$f$ spin orbit components, which cannot be accounted for by an  increase of $\alpha$. 
         ~A good fitting is obtained if one extra component (green line, designated as Au$_{\rm A2}$) is considered that have similar width as the Au bulk substrate component (red line, designated as Au$_{\rm B}$). This peak appears at 0.4~eV higher BE from  Au$_{\rm B}$, in agreement with our earlier work.\cite{Sadhukhan_archive} 
  ~For $t_d$= 4 min, the asymmetry is enhanced and one more component (blue line, designated as Au$_{\rm A1}$) is required to get a good fit. Au$_{\rm A1}$ and Au$_{\rm A2}$ appear at 0.75 and 0.4~eV higher BE from  Au$_{\rm B}$, respectively [Fig.~\ref{Au4fcorefit}(b)].  For 8-12 min Sn deposition, the asymmetry is  further enhanced with a visible discontinuity in the slope at 85~eV [Fig.~\ref{Au4fcorefit}(c,d)]. 
  ~Both Au$_{\rm A1}$ and Au$_{\rm A2}$ increase in intensity at the expense of Au$_{\rm B}$ [Fig.~\ref{Au4fcorefit}(c, d)]. 
~For $t_d$= 15 min, all the components are of similar intensity [Fig.~\ref{Au4fcorefit}(e)].
         ~With further increase in $t_d$, both Au$_{\rm A2}$ and Au$_{\rm B}$ decrease continuously in intensity, while the Au$_{\rm A1}$  becomes predominant [Fig.~\ref{Au4fcorefit}(e-h)]. The different parameters obtained from the curve fitting of the Au 4$f$ spectra are shown in Table~S.I of the SM.\cite{supplement}

          In case of Sn 4$d$, the wide peak observed for Sn deposition in the range of $t_{d}$= 2-8 min, cannot be explained by a single component. The presence of two components for $t_{d}$= 2 min has also been established earlier from higher resolution UPS spectrum.\cite{Sadhukhan_archive} 
          ~For $t_{d}$= 4 min, two distinct components,  Sn$_{\rm A1}$ (blue line) and Sn$_{\rm A2}$ (green line), are observed at 24.4~eV and 24~eV [Fig.~\ref{Sn4dcorefit}(a)] corresponding to Sn 4$d_{5/2}$ having similar linewidth as Sn metal. In Fig.~\ref{Sn4dcorefit}(b,c)  the relative intensity of Sn$_{\rm A2}$ component monotonically decreases compared to Sn$_{\rm A1}$ with increasing $t_d$. The Sn 4$d$ spectra is dominated by Sn$_{\rm A1}$, which slightly shifts ($\approx$ 0.2~eV) towards higher BE  for $t_d$= 8 min. This peak position remains unchanged up to $t_d$= 57 min [also see Fig.~\ref{core_cov}(b)]. Similar to the \Autwo, 
          ~ Sn$_{\rm A2}$ becomes negligible  and only Sn$_{\rm A1}$ component exists up to the highest coverage [Fig.~\ref{Sn4dcorefit}(d)]. The different parameters obtained from the curve fitting of Sn 4$d$ spectra are shown in Table~S.II of the SM.\cite{supplement}

         The relative intensity variation of three Au 4$f$ components mentioned above is shown in Fig.~\ref{intensities}(a).  Au$_{\rm B}$  exhibits an initial  rapid decrease up to $t_{d}$ $\approx$20 min, and then gradually decreases,
         ~and this is further  discussed later in relation to the determination of the layer thickness.  In  contrast to Au$_{\rm B}$, Au$_{\rm A1}$ increases with $t_d$, but becomes almost constant for $t_d$$\geq$ $\approx$39 min. The intensity variation of Au$_{\rm A1}$ indicates that it is related to Au atoms that diffuse through the Sn layer forming the Au-Sn layer. 
         ~Moreover, its position is $\approx$0.9~eV shifted towards higher BE compared to Au$_{\rm B}$ and  remains almost unchanged throughout the deposition range (Table S.I of SM)\cite{supplement}  indicating the intermixing of Au and Sn. 
         The component Au$_{\rm A2}$ initially increases   up to $t_d$= 12 min and exhibits a broad hump for  $t_d$= 12-18 min, and eventually  decreases to a small value. 
         ~Its BE  remains almost unchanged (Table S.I of SM).\cite{supplement} This  intensity variations of Au$_{\rm A2}$  suggests that it is related to the interface region, since due to $\approx$ 2~nm mean-free path in XPS, the signal from interface region decreases with increasing Au-Sn layer thickness. However, for small Sn coverages, for example, $t_d$= 2 min, Au$_{\rm A2}$ would represent the Au-Sn alloy.\cite{Sadhukhan_archive} 
         
         ~ The Sn 4$d$ components also exhibit similar intensity variation as those of Au 4$f$ [Fig.~\ref{intensities}(b)]. The intensity of the  Sn$_{\rm A1}$ component increases continuously with $t_d$ and becomes nearly constant for larger $t_d$ ($\geq$ $\approx$39 min). 
         	~In contrast, the intensity of Sn$_{\rm A2}$ increases initially showing a maximum at $t_d$= 12 min and decreases gradually and eventually becomes very small. 
         	~The characteristic intensity variations of the Sn 4$d$ components suggest that Sn$_{\rm A2}$ is  possibly related to the interface region, while Sn$_{\rm A1}$ is related to the Au-Sn layer.
            The intensity variation of total Au 4$f$ and Sn 4$d$ (adding all components) are plotted in Fig.~\ref{intensities}(c) after dividing by their respective photoemission cross-sections.\cite{Yeh85} The total Au 4$f$ (Sn 4$d$) intensity initially decreases (increases) and eventually becomes nearly constant, 
            ~indicating formation of the thick  Au-Sn layer with negligible signal from the Au substrate or the interface.  

\underline{{\bf Determination of the composition and thickness:}}   We ascertain the average composition of the whole Au-Sn layer by considering the area of (Au$_{\rm A1}$+ Au$_{\rm  A2}$) to represent the Au signal (Au$_{ \rm B}$ is not considered since it is from the substrate), while area of  (Sn$_{\rm A1}$+ Sn$_{\rm A2}$) represents the Sn signal. These areas are divided by their respective photoemission cross-sections and the composition averaged over different $t_d$ turns out to be  Au$_{1.1}$Sn ($i.e.$ $\approx$ AuSn).
 We can also find the composition of the layer and the interface separately, and this turns out to be AuSn and Au$_{1.8}$Sn, respectively. This indicates that the interface is more Au rich compared to the layer. 
 ~AuSn in bulk form  has been studied earlier by XPS\cite{Attekum79,Taylor95,Friedman73}, and the BE of Au 4$f$ and Sn 4$d$ in bulk AuSn agree with what we observe here for Au$_{\rm A1}$ and Sn$_{\rm A1}$. This further supports our proposition that  AuSn forms as a result of  Sn deposition  on Au(111) at RT. 
Note that the variation of Au$_{\rm B}$ in Fig.~\ref{intensities}(a) exhibits an excellent fit with an exponential curve (dashed line) indicating that AuSn grows as a layer  on the Au substrate. The evidence of layered growth is also obtained from STM for deposition of about a monolayer of Sn at RT (Fig.~S1).\cite{supplement}
	
	 We have estimated the thickness of the AuSn layer  using the well known expression $I_S$= $I_{S\infty}$$\times$$e$$^{(-d/\lambda)}$ where $I_S$ is Au$_{\rm B}$ intensity, $I_{S\infty}$ is Au$_{\rm B}$ intensity with no Sn deposition, $d$ is the thickness and $\lambda$ is the inelastic mean free path. For AlK$\alpha$ radiation, $\lambda$ of AuSn at Au 4$f$ kinetic energy is 2.18 nm.\cite{tpp2m} Taking this value of $\lambda$, we calculate 
~	$d$ and find that it varies  linearly with $t_d$, as shown by a straight line fit in Fig.~S3.\cite{supplement} Thus, for $t_d$= 15 and 45 min, $d$ is 3.3 and 8.9 nm, respectively. For $t_d$= 57 min,  $I_S$ $i.e.$ Au$_{\rm B}$ is negligible  [Fig.~\ref{intensities}(a)],  so we have extrapolated $d(t_d)$ to obtain $d$= 11.4 nm for $t_d$= 57 min. It may be noted that the corresponding values of $d$  estimated from the LEED based method discussed in Section II are 2.8, 8.4 and 10.7 nm for $t_d$= 15, 45 and 57 min, respectively. Thus, the estimate of $d$ using XPS and LEED methods are in  agreement with a maximum error of about 10\%.  

    \begin{figure}[htb]
    	\vskip -5 mm
    	\includegraphics[width=120mm]{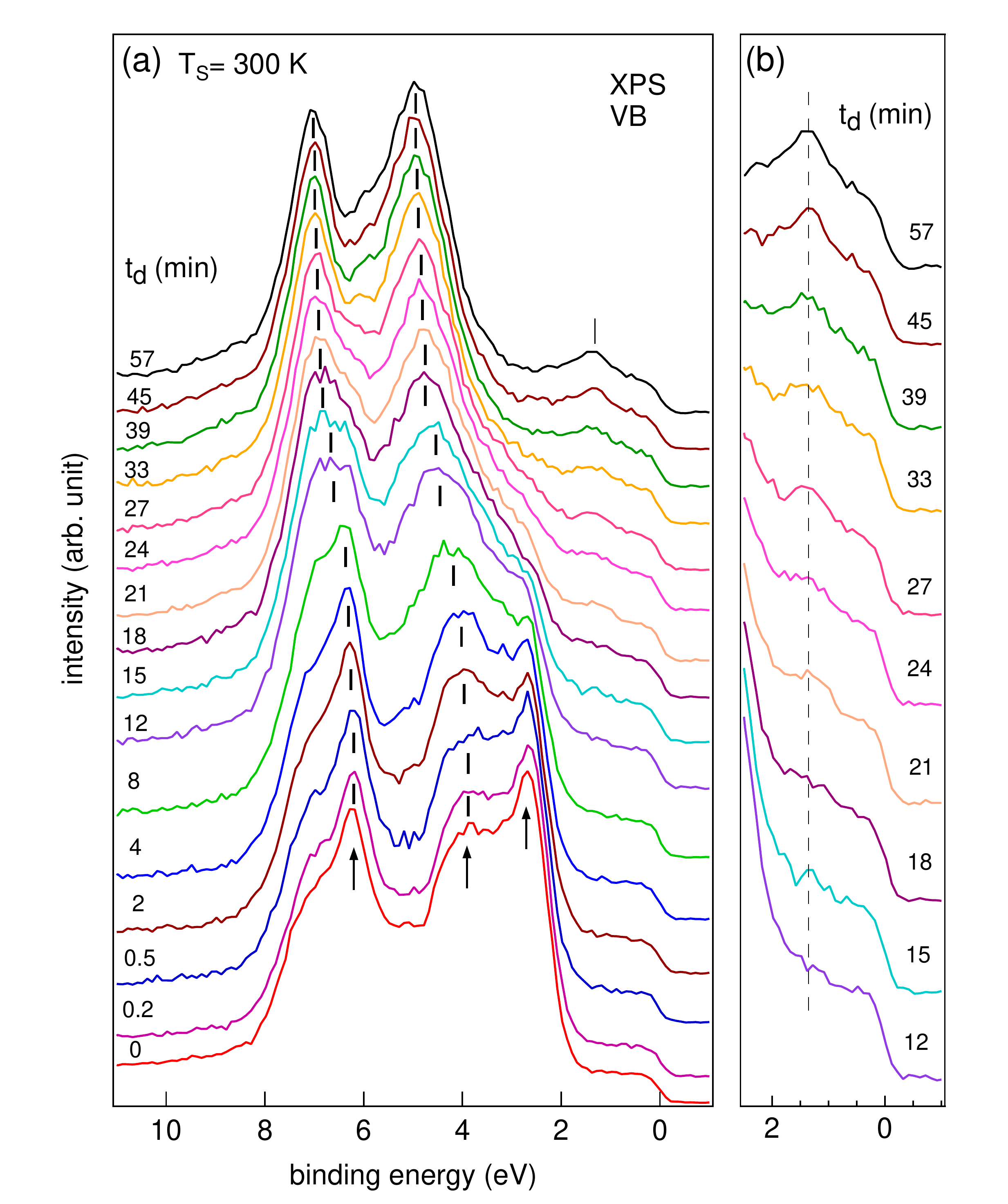} 
    	\caption{(a) The valence band spectra of AuSn as a function of  $t_d$ at  RT  using monochromatic AlK$_{\alpha}$ radiation. The spectra are normalized to the same height at the maximum and staggered along the vertical axis, the zero of each spectrum is given by the flat region above the Fermi level ($E_F$ at 0 eV). (b) The region below $E_F$ is shown in an expanded scale to highlight the weak peak at 1.3~eV, see the dashed black vertical line.} 
    	\label{VB_cov}
    \end{figure}        
     
\subsection{XPS valence band of AuSn: experiment and theory}
Fig.~\ref{VB_cov}(a) presents XPS valence band of AuSn grown on Au(111) as a function of $t_{d}$ at  RT. The  Au metal VB exhibits prominent $d$-like states with a sharp peak at 2.7~eV, a broad hump at 3.9~eV and another peak at 6.2~eV (see black arrows), while the $s$-like states  appear between 
~2~eV and $E_F$. As Sn is deposited, the Au 5$d$ peak at 2.7~eV  is clearly suppressed, and both the peaks at 3.9~eV and 6.2~eV shift towards higher BE for $t_d$$\leq$21 min (shown by black ticks). For $t_d$$\geq$21 min, where the substrate Au VB signal is almost fully suppressed [as also shown by negligible Au$_{\rm B}$ component in Fig.~\ref{intensities}(a)], the VB  would represent AuSn and consequently its shape does not change as the peaks do not shift any more from  4.9~eV and 7~eV, respectively (black ticks). 
~It is interesting to note the emergence of a peak  at 1.3~eV  for $t_d$$\geq$21 min and it becomes progressively prominent with $t_d$  [Fig.~\ref{VB_cov}(b)].   

\begin{figure*}[tb]
	\centering
	\vskip -20mm
	\includegraphics[width=160mm]{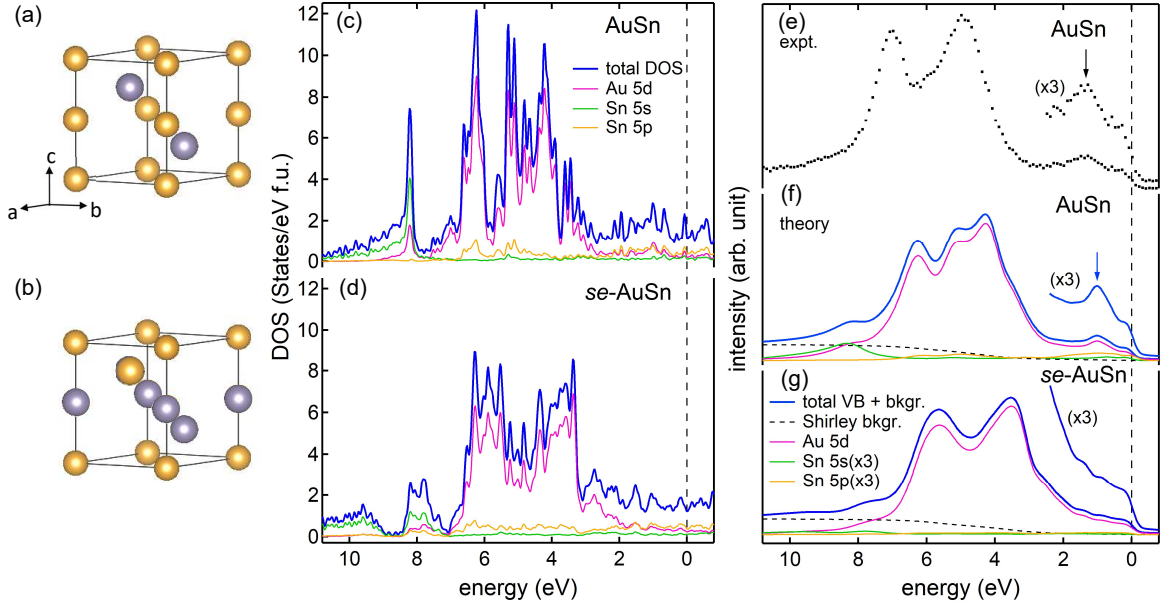} 
	\vskip -25mm
	\caption{ The crystal structure of (a) AuSn and (b) site-exchanged AuSn (se-AuSn),  where yellow and grey filled circles denote the Au and Sn atoms, respectively. 
		~ The calculated density of states (DOS) and PDOS of (c) AuSn and  (d) se-AuSn. (e) The experimental XPS VB of AuSn obtained by $t_d$= 57 min deposition at RT  compared with the calculated VB  of (f)  AuSn and (g) se-AuSn. } 
	\label{AuSn_DOS}
\end{figure*}


      In order to interpret the XPS valence band spectra in Fig.~\ref{VB_cov} and the peak at 1.3~eV in particular, we have performed DFT calculation for AuSn
      ~using the bulk structure from literature [Fig.~\ref{AuSn_DOS}(a)].\cite{Jan63} 
      ~The total DOS of AuSn in Fig.~\ref{AuSn_DOS}(c) as well as the calculated VB in  Fig.~\ref{AuSn_DOS}(f)  show a peak around 1 eV  (blue arrow) that resembles the  peak at 1.3 eV (black arrow) in the experimental XPS VB for $t_d$= 57 min, where any contribution from the substrate is absent  [Fig.~\ref{AuSn_DOS}(e)]. The PDOS shows that it  arises primarily from  Au 5$d$ states, with some admixture of  Sn 5$p$ states. Since the photoemission cross-section of the Au 5$d$  is much larger than that of the Sn states (see Section II), a clear peak is observed in the calculated VB. The 3-7~eV region is also dominated by the Au 5$d$ states, whereas the sharp peak at 8.2~eV has main contribution from Sn 5$s$ states with some admixture of Au 5$d$ states.
~    In the calculated VB, the  peaks at 4.3 and 6.3 eV are  primarily related to the Au 5$d$ states  and  are in good agreement with the peaks at 4.9 and 7 eV in the experimental VB [Fig.~\ref{AuSn_DOS}(f)]. Even the shoulder at 5.9 eV in between these two peaks has its counter part in the theory at 5.2 eV, although its intensity is overestimated in the latter.  

We have performed DFT calculations for site-exchanged AuSn (se-AuSn) [Fig.~\ref{AuSn_DOS}(b)] by interchanging one Au atom in the 2a position with a Sn atom at the 2c position  to probe the influence of anti-site defects on the electronic structure of AuSn. The peak at 1 eV observed in Fig.~\ref{AuSn_DOS}(c,f) for AuSn, is hardly visible in this case 
~in  the calculated VB  [Fig.~\ref{AuSn_DOS}(g)]. 
~
~A close inspection of the se-AuSn DOS in Fig.~\ref{AuSn_DOS}(d) shows that although there is a weak peak around 1 eV, it is dominated by the Sn 5$p$ states that have low photoemission cross-section. Moreover, the peaks at 3.5 and 5.7 eV  in Fig.~\ref{AuSn_DOS}(g) are in worse agreement [compared to AuSn in Fig.~\ref{AuSn_DOS}(f)]  with the experimental peaks at 4.9 and 7 eV in Fig.~\ref{AuSn_DOS}(e). 
~Furthermore, the shoulder at 5.9 eV in the experiment is not observed for se-AuSn. 
~On the contrary, the satisfactory agreement  of the experiment with the theoretical VB spectrum of AuSn (without site exchange) in Figs.~\ref{AuSn_DOS}(e,f) suggests that anti-site defects do not play a dominant role and the AuSn layer is by and large an ordered intermetallic compound.

        \subsection{Core-level XPS of Au-Sn phases formed by high temperature treatment}

        \begin{figure}[tb]
        	\centering
        	\includegraphics[width=120mm]{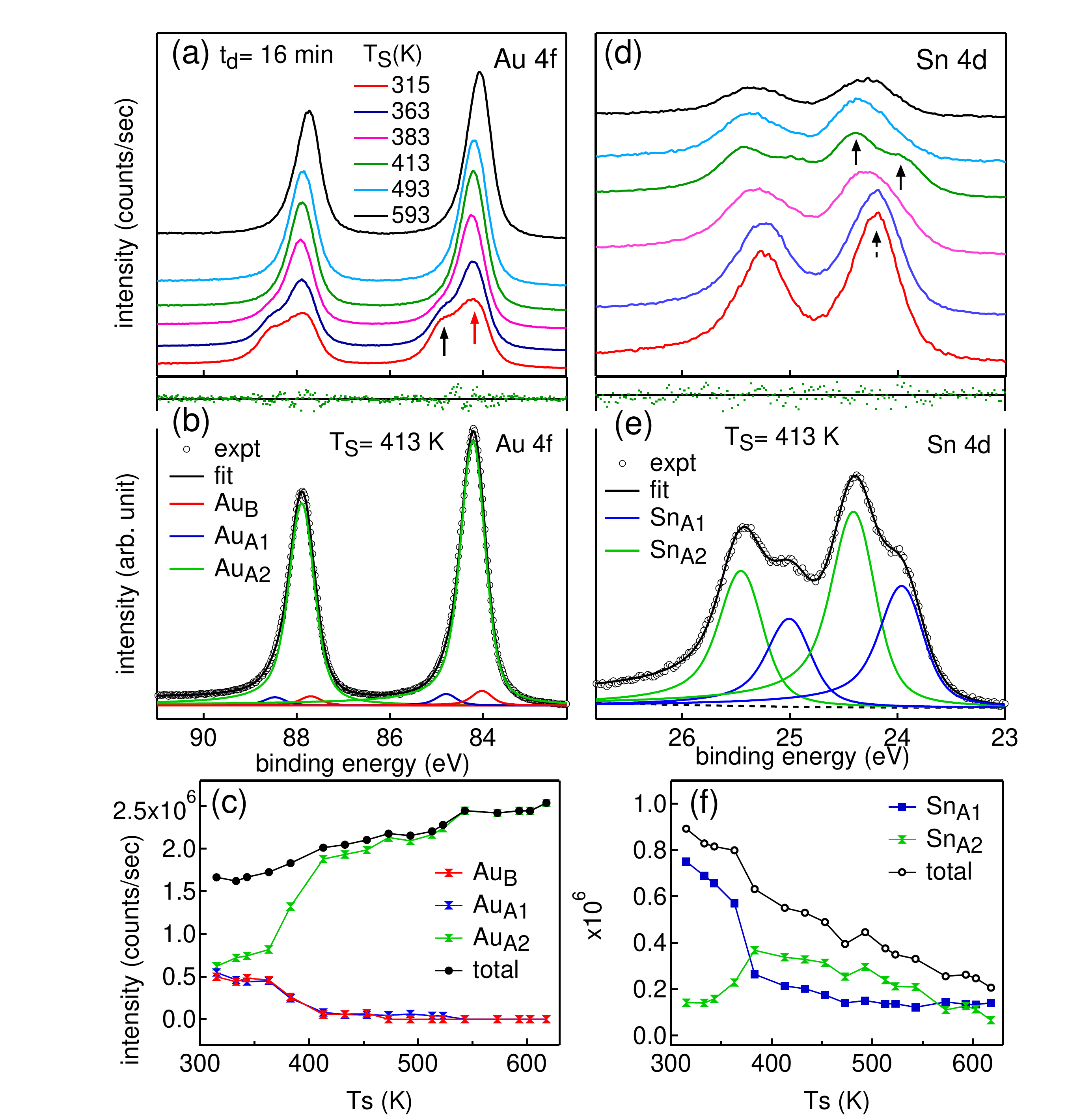} 
        	\caption{XPS core level spectra for $t_d$= 16 min Sn deposition for (a) Au 4$f$ and (d) Sn 4$d$ as a function of  substrate temperature during deposition  ($T_S$). 
        		~A representative fitting for $T_S$= 413 K  (black line) of (b) Au 4$f$ and (e) Sn 4$d$ core level spectra (black open circles) that are normalized to the same height along with the different components.  	Intensities of the different components and total  (c) Au 4$f$ and (f) Sn 4$d$ obtained from the least square fitting  after dividing by their respective photoemission cross-sections.}
        	\label{core_Ts}
        \end{figure}

\begin{figure}[htb]
        	\centering
        	\includegraphics[width=120mm]{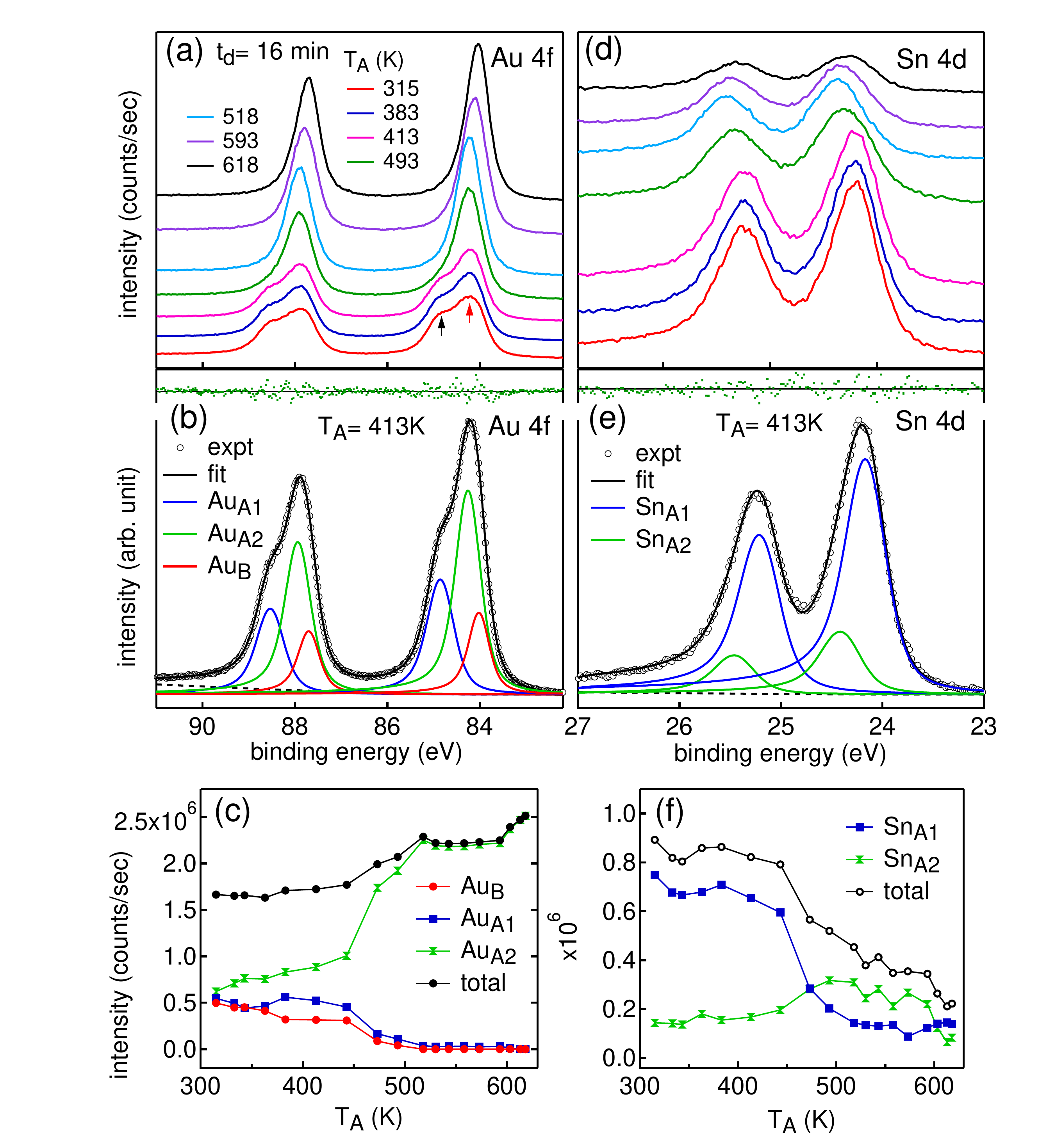}
        	\caption{XPS core level spectra for $t_d$= 16 min Sn deposition for (a) Au 4$f$ and (d) Sn 4$d$ as a function of post annealing temperature ($T_A$). 
        		A representative fitting for $T_A$= 413 K (black line) of (b) Au 4$f$ and (e) Sn 4$d$ core level spectra (black open circles) that are normalized to the same height along with the different components.  	Intensities of the different components and the total  (c) Au 4$f$ and (f) Sn 4$d$ obtained from the least square fitting  after dividing by their respective photoemission cross-sections.}
        	
        	\label{core_Ta}
        \end{figure}

         In this section, we  discuss the effect of high temperature deposition for $T_S$ up to 618 K on  Au-Sn alloy formation [Fig.\ref{core_Ts}]. For each $T_S$, a fresh deposition of Sn on clean Au(111) for $t_d$= 16 min was done. This is compared with the effect of step-wise post-annealing  of the AuSn layer up to high temperatures  ($T_A$\,$\leq$\,618 K). The AuSn layer was grown  by  Sn deposition for $t_d$= 16 min at RT  [Fig.\ref{core_Ta}]. The heat treatment parameters for the compositions directly discussed in the paper are shown in Table.~\ref{heattreatment}, while in Tables S.III and S.IV of SM\cite{supplement} the heat treatment parameters for  all the compositions 
	~are shown. 

        \begin{table}  
        	\caption{Au-Sn preparation parameters by high temperature treatment and the corresponding  compositions obtained from  XPS. The  Knudsen cell temperature (1140$\pm$5 K), the normal incidence geometry and distance of the sample with respect to the cell are unchanged. The measurements have been performed  at RT in all cases. Each entry in the $t_d$ column indicates a deposition on  sputter-annealed clean Au surface. The layers for which ARPES has been performed are highlighted in bold and their LEED patterns are indicated.}
        	\begin{tabular}{|c|c|c|c|} \hline 
        		\multicolumn{3} {|c|}  {Preparation parameters} &  \multicolumn{1} {|c|}  {Composition} \\  
        		t$_d$ (min) &  T$_S$ (K)  & T$_A$ (K) & \\ \hline  
        		16 & 315 &  & Au$_{1.1}$Sn$^{\dagger}$   \\ 
        		16 & 383 &  & Au$_{2.4}$Sn \\
        		{\bf 16} & {\bf 413} &  &  {\bf Au$_{3.6}$Sn,  p(3$\times$3)R15$^{\circ}$} \\
        		16 & 493 &  & Au$_{4.8}$Sn  \\
        		16 & 593 &  & Au$_{9.3}$Sn \\
        		16 & 618 &  & Au$_{12.3}$Sn \\
        		 \hline
        		16 &  & 315 & Au$_{1.1}$Sn$^{\dagger}$  \\
        		&  & {\bf 383} &  {\bf Au$_{1.6}$Sn, mixed}\\
        		&  & {\bf 443} &  {\bf Au$_{1.9}$Sn, mixed} \\
        		&  & 493 & Au$_{4}$Sn \\
        		&  & {\bf 518} & {\bf Au$_{5.1}$Sn ($\sqrt{3}\times\sqrt{3}$)R30$^{\circ}$}  \\
        		&  & 593 & Au$_{6.5}$Sn   \\ 
        		&  & 618 & Au$_{11.3}$Sn   \\ 
        		\hline

        	\end{tabular}
        	
        	\noindent$^{\dagger}$average over different $t_d$. 
        	\label{heattreatment}
        \end{table}
    

 {\bf \underline{High temperature deposition:}}          As a function of $T_S$, the shape and position of  Au 4$f$ spectra change substantially [Fig.~\ref{core_Ts}(a)]. 
          ~The  bulk component Au$_{\rm B}$ is  suppressed in intensity with increasing $T_{S}$ [Fig.~\ref{core_Ts}(b,c)].  Here, the AuSn layer related Au$_{\rm A1}$ component (black arrow)  at 84.9~eV decreases in intensity [Fig.~\ref{core_Ts}(c)]. 
          ~On the other hand, the \Autwo component (red arrow) that is related to a Au rich phase at the interface increases in intensity. Thus, the  variation of \Auone and \Autwo here is different from that with $t_d$ at RT shown in Fig.~\ref{intensities}(a). 
          Sn 4$d$ spectrum also shows considerable change in the line shape with $T_S$, as shown by the black arrows in Fig.~\ref{core_Ts}(d). 
          The two components Sn$_{\rm A1}$   and Sn$_{\rm A2}$ are related to the AuSn alloy layer and the interface, respectively [Fig.~\ref{core_Ts}(e,f)] and at $T_A$= 315~K,  Sn$_{\rm A1}$ is dominant, as also shown in Fig.\ref{intensities}(b). However, with increase of $T_S$,  \Snone  decreases and \Sntwo increases, thus showing similar variation as Au 4$f$ spectra shown in Fig.~\ref{core_Ts}(c) and opposite trend compared to the variation with $t_d$ at RT shown in Fig.~\ref{intensities}(b).
          The compositions, determined following the same method as in Section III.A, show that  the layer becomes  enriched in Au (or diluted in Sn) with increasing $T_S$ (Table \ref{heattreatment}) reaching Au$_{12.3}$Sn at $T_S$= 618 K.

\begin{figure}[htb]
        	\centering
        	\includegraphics[width=140mm]{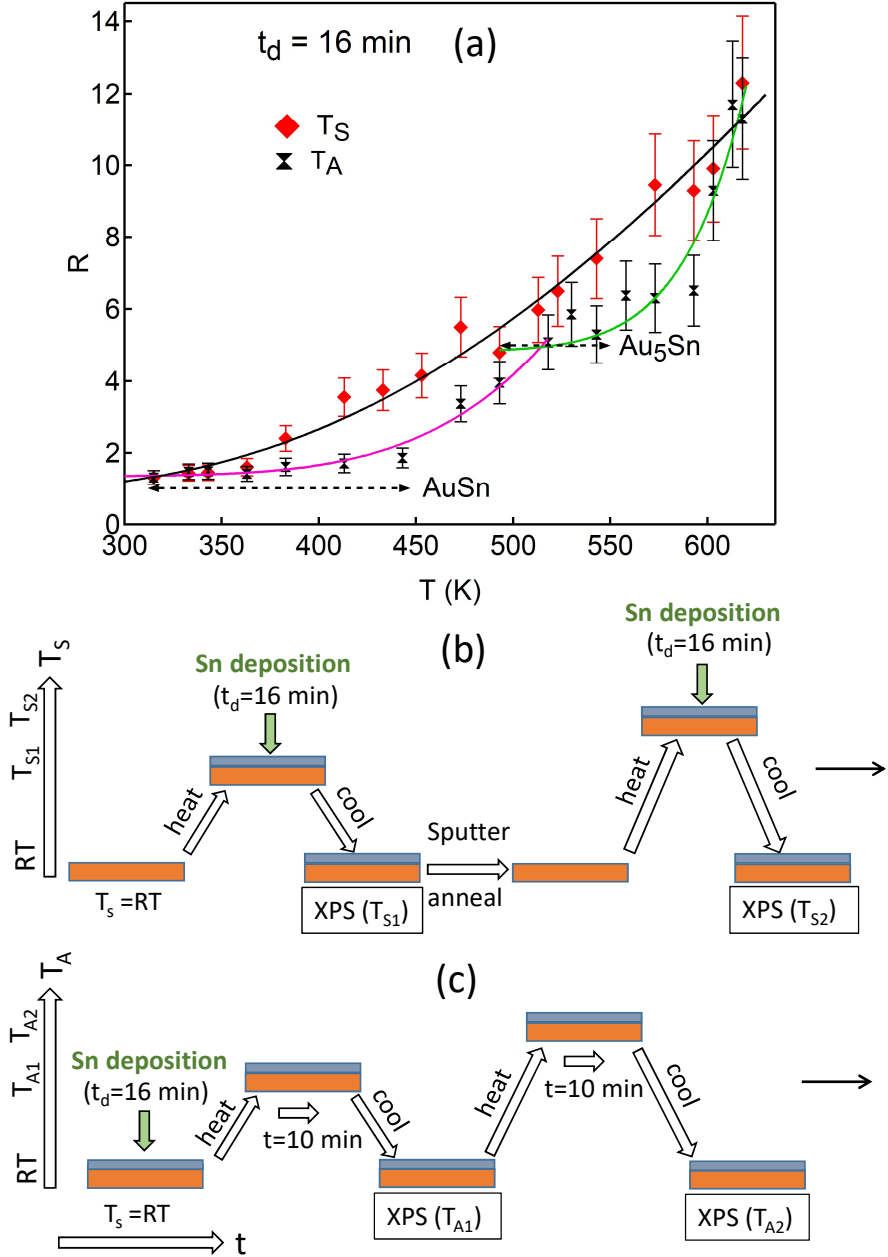}
        	\vskip -8mm
        	\caption{  (a) The composition variation of the Au-Sn film depicted by $R$ (= Au:Sn ratio) as a function of temperature ($T_A$ and $T_S$). Least-square exponential fit to the data are shown by the solid  curves (see text). A schematic diagram of  (b) high temperature deposition as function of  $T_S$ at $T_{S1}$ and $T_{S2}$ and (c) post annealing as a function of $T_A$ at  $T_{A1}$ and $T_{A2}$. The Au-Sn layer is indicated by grey, while the  Au(111) substrate is indicated by orange color.  Vertical and horizontal axes denote temperature and time ($t$), respectively. The horizontal black arrows on the right hand side denote continuation of the process to higher temperatures. }
        	\label{compratio}
        \end{figure}

{\bf \underline {Post annealing:}}        The effect of post annealing by increasing $T_A$ on   Au 4$f$ and Sn 4$d$ spectra of the AuSn layer 
        ~is shown in Fig.~\ref{core_Ta}. 
        ~The behavior of the core-level line shapes and the variation of the intensities of the different components are similar to that of $T_S$ (Fig.~\ref{core_Ts}). 
        ~The Au$_{\rm A1}$ component  at 84.9~eV that represents the AuSn layer decreases in intensity [Fig.~\ref{core_Ta}(c)]. 
        ~Notably, 
        ~the  composition obtained by post annealing $e.g.$  Au$_{6.5}$Sn for $T_A$= 593~K is considerably  less Au rich compared to high temperature deposition $e.g.$   Au$_{9.3}$Sn at $T_S$= 593~K, the reasons for this difference are discussed later (Fig.~\ref{compratio}). Another point to note is that  the plasmon related loss features of Sn 4$d$ are not observed  after the heat treatment with both $T_S$ as well as $T_A$ providing evidence for absence of a Sn only top layer [Fig.~S2(b,c)].\cite{supplement} 
        
{\bf\underline {Comparison of the high temperature deposition and post-annealing methods:}}          An interesting trend that is similar for both $T_S$ and $T_A$ heat treatments is the suppression of the substrate related \AuB component [Figs.~\ref{core_Ts}(c), \ref{core_Ta}(c)]. This clearly  indicates that the thickness  of the alloy layer increases as it becomes diluted in Sn or enriched in Au with Au:Sn ratio ($R$) increasing with heat treatment.   
                  ~An estimate of the  thickness ($d$) of  starting AuSn layer at RT is 3.3 nm obtained from the exponential decay of \AuB (Fig.~S3).\cite{supplement} Considering the same exponential decay with $\lambda$= 2.18 nm, $d$ turns out to be 8.1 nm at $T_S$= 413~K, whereas it is 4.3 nm at $T_A$= 413~K. For  temperatures above 493~K, 
            \AuB is almost zero indicating that  $d$ is $\geq$10~nm (Fig.~S3).\cite{supplement}      
                  ~The increase in thickness would result in the increase of $R$ ($i.e.$ Au enrichment or dilution of Sn), since the amount of Sn is unchanged because no further deposition or desorption occurs.  
        
        The increase of $R$ with heat treatment and the concomitant decrease of the AuSn layer related \Auone and \Snone signals and increase of the interface related \Autwo and \Sntwo signals indicate inter-diffusion of Au and Sn at the interface, which grows in thickness on both sides and becomes diluted (enriched)  in Sn(Au). Enrichment of Au due to its out-diffusion into Sn and increase in the width of this region leads to the increase of the total Au signal [Fig.~\ref{core_Ts}(c),\ref{core_Ta}(c)]. On the other hand, Sn becomes dilute as it diffuses deeper into Au to the order of 10~nm  ($i.e.$ the thickness of the Au-Sn layer). Thus, its  signal decreases [Fig.~\ref{core_Ts}(f),\ref{core_Ta}(f)] due to limited probing depth of XPS ($\lambda$$\approx$2~nm). Such diffusion in  Au leading to disappearance of the Sn signal has been reported for Sn oxides above 800 K and no thermal desorption was observed.\cite{Zhang94} 

     An interesting difference between high temperature deposition ($T_S$) and post annealing ($T_A$) methods lies in the variation of $R$ as a function of temperature ($T$): 
      $R$($T$) is larger in the former for the whole $T$ range [Fig.~\ref{compratio}(a)]. 
        ~For  heat treatment by $T_A$, AuSn ($R$= 1), which is a  bulk compound, is already formed by $t_d$= 16 min deposition at RT. In Fig.~\ref{compratio}(a),   up to about 450 K, $R$ barely changes  but increases thereafter. 
        ~It is interesting  to note  that above $R$= 1, the next bulk compound is reported for  $R$= 5 $i.e.$  Au$_5$Sn.\cite{Osada74} 
         In Fig.~\ref{compratio}(a), the $R$ values seem to stabilize around 5 in the range between 500- 550 K above which it  increases quite steeply. 
         ~Thus, for each of the stable bulk compositions ($R$= 1 and 5), there is a plateau over a finite temperature range (dashed black  lines with arrows at both ends) followed by an increase, indicating existence of an activation barrier.  In order to determine its nature, we note that the diffusivity of Au in Sn increases exponentially with $T$.\cite{Dyson66} So, we have fitted  $R(T_A)$ by an exponential function $d_0$$\times$exp$^{-(Q/kT)}$ in the range of 315 K to 518 K, where $Q$ is the activation barrier in kcal/mol, $k$ is the Boltzmann constant in kcal\,mol$^{-1}$K$^{-1}$, $d_0$ is a pre-exponential factor and a constant offset is also considered. 
         ~The fitting is reasonable  (pink curve) and the activation barrier for transformation of AuSn to Au$_5$Sn turns out to be $Q$= 9 kcal/mol.  A fitting  using the same expression in the range 493 K to 618 K  (green curve) gives $Q$= 24 kcal/mol for transformation of Au$_5$Sn to  further Au rich compositions $e.g.$  Au$_{10}$Sn that is also a stable bulk phase.\cite{Torleif11}  It is expected that  $R$ might increase to even higher values with temperature since inter-diffusion of Au and Sn would continue. The maximum $R$ observed in this work is 12.3 for $T_A$= 618 K.   
            
        Unlike $R(T_A)$,   $R(T_S)$  does not show any plateaus corresponding to the stable bulk compositions [Fig.~\ref{compratio}(a)]. Thus,  fitting the whole temperature range (315$\leq$T$\leq$618) with a single exponential function  provides reasonably good fit (black line)  with $Q$= 4 kcal/mol, which is substantially smaller than  the reported  $Q$ of 11- 17.7 kcal/mol for diffusion of  Au in bulk Sn single crystal.\cite{Dyson66}
        
        The difference between $R(T_A)$ and $R(T_S)$ stems from the fact that these are quite different processes of heat treatment, as shown in Fig.~\ref{compratio}(b,c). 
		The main difference is that for $R(T_S)$, the deposition is performed at successively  higher temperatures for $t_d$= 16 mins each, whereas for $R(T_A)$ one deposition is done at $t_d$= 16 mins and it is post annealed successively to high temperatures for 10 mins at each temperature. After each heat treatment, the sample is cooled to RT for XPS measurements in both cases.
		Thus, for $R(T_S)$, besides the effect of substrate temperature, the diffusivity of the Sn atoms  would be further enhanced  because these are  impinging the surface with sizable  kinetic energy acquired by thermal evaporation from the K-cell operating at 1140~K.	Thus, Au-Sn inter-diffusion is likely to be more favored in this case compared to $T_A$ where the kinetic energy factor is absent. For this reason, $Q$ has a lower value ($Q$= 4 kcal/mol) for $R(T_S)$ compared to $R(T_A)$, although the annealing time during deposition is much less for $R(T_S)$ ($t_d$= 16 min), whereas for $R(T_A)$ the same layer is cumulatively annealed at different $T_A$  [Fig~\ref{compratio}(b,c)].
		
		A decrease of the sticking coefficient of Sn  with increasing $T_S$ could also play a role in the high temperature deposition.  
		~Since decrease of the  sticking coefficient would mean lesser amount of Sn on the Au surface, we have simulated its effect on $R$ by decreasing  $t_d$ instead. 
		~Thus, since in the above experiments we have used $t$= 16 min (Table~I),  $t_d$$<$16 min would 
		~simulate the effect lower sticking coefficient on $R$. From seven measurements in the range 10 min$\leq$$t_d$$\leq$20 min, we find that $R$ does not vary significantly  (see Table S.V of SM).\cite{supplement} This shows that a change of the sticking coefficient is not the cause of the above discussed difference between $R(T_S)$ and $R(T_A)$.
		Desorption with increasing $T_A$ could play a role in post annealing. But, absence of Sn desorption in the studied $T_A$ range up to 618 K has been confirmed by the  absence of  Sn signal in the mass spectrometer during annealing.

 \subsection{XPS valence band of the Au rich Au-Sn phases}

 \begin{figure}[tb]
 	\vskip -5 mm
 	\includegraphics[width=165mm]{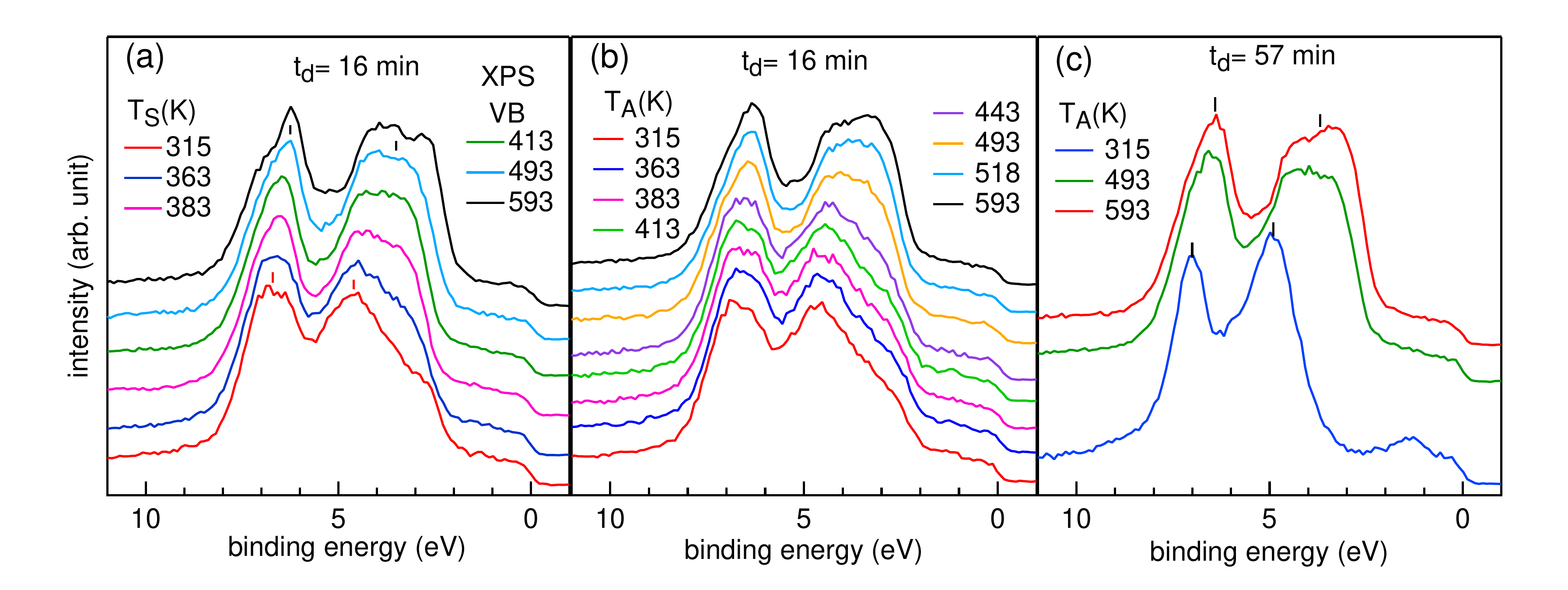} 
 	\caption{ (a) The effect of substrate temperature $T_S$ on the XPS VB spectra of AuSn  obtained with $t_d$= 16 min Sn deposition. The effect  of post annealing  as a function of $T_{A}$  for  (b) $t_d$= 16 min and (c) $t_d$= 57 min Sn deposition at RT.   All the spectra are normalized to the same height and staggered along the vertical axis for clarity of presentation. } 
 	\label{VB_Ts}
 \end{figure}

\underline{{\bf Effect of high temperature treatment on the XPS Valence band of Au-Sn:}} The XPS valence band spectra show systematic changes in its shape as $R$ increases with high temperature treatment  [Fig.~\ref{VB_Ts}]. 
 ~At $T_{S}$= 315\,K, the VB shows two peaks around 6.7~eV and 4.6~eV marked as red ticks in Fig.~\ref{VB_Ts}(a). On the other hand, the states  in the BE range of 2.5 to 4.6~eV are suppressed $w.r.t.$  Au metal VB. With increasing $T_S$, the states in this region increase in intensity and both the marked peaks shift towards lower BE. For example, at $T_{S}$= 493~K, a broad peak centered at 3.5~eV is observed along with the higher BE peak at 6.5~eV (black ticks).
  Similar behavior of the VB  is observed with $T_A$ [Fig.~\ref{VB_Ts}(b)], but, as discussed in the previous section, higher $T_A$ is required to attain a particular $R$ and hence a particular VB shape, in agreement with Fig.\ref{compratio}(a). 
~   For $t_d$= 57 min, annealing to  $T_A$= 493 and 593~K widens and shifts the lower BE peak to 3.7~eV, while the higher BE peak at 7~eV shifts to 6.4~eV [see black ticks in Fig.~\ref{VB_Ts}(c)]. The weak  peak at 1.3~eV at RT that is related to Au 5$d$-Sn 5$p$ hybridized states of AuSn [discussed earlier, see Fig.~\ref{AuSn_DOS}(c)] is not observed for  $T_A$= 493- 593~K due to change in the composition. As discussed in section III.B, the shape of the AuSn XPS VB at RT is different for $t_d$= 57 min compared to $t_d$= 16 min because there is considerable substrate Au VB contribution in the latter (Fig.\ref{VB_cov}). However, with heat treatment $e.g$ $T_A$= 493~K, since the Au-Sn layer becomes thicker,  almost no contribution of bulk Au is observed for $t_d$= 16 min. This is also evident from the similarity of the VB between $t_d$= 16 min [Fig.~\ref{VB_Ts}(b)] and $t_d$= 57 min [Fig.~\ref{VB_Ts}(c)] for  $T_A$$\geq$493~K. This is also consistent with the core level analysis of Au 4$f$ in Figs.~\ref{core_Ts}(c), \ref{core_Ta}(c)  that shows absence of the Au$_{\rm B}$ component for $T_A$$\geq$493~K.

 \begin{figure}[tb]
 	\centering
 	\includegraphics[width=180mm]{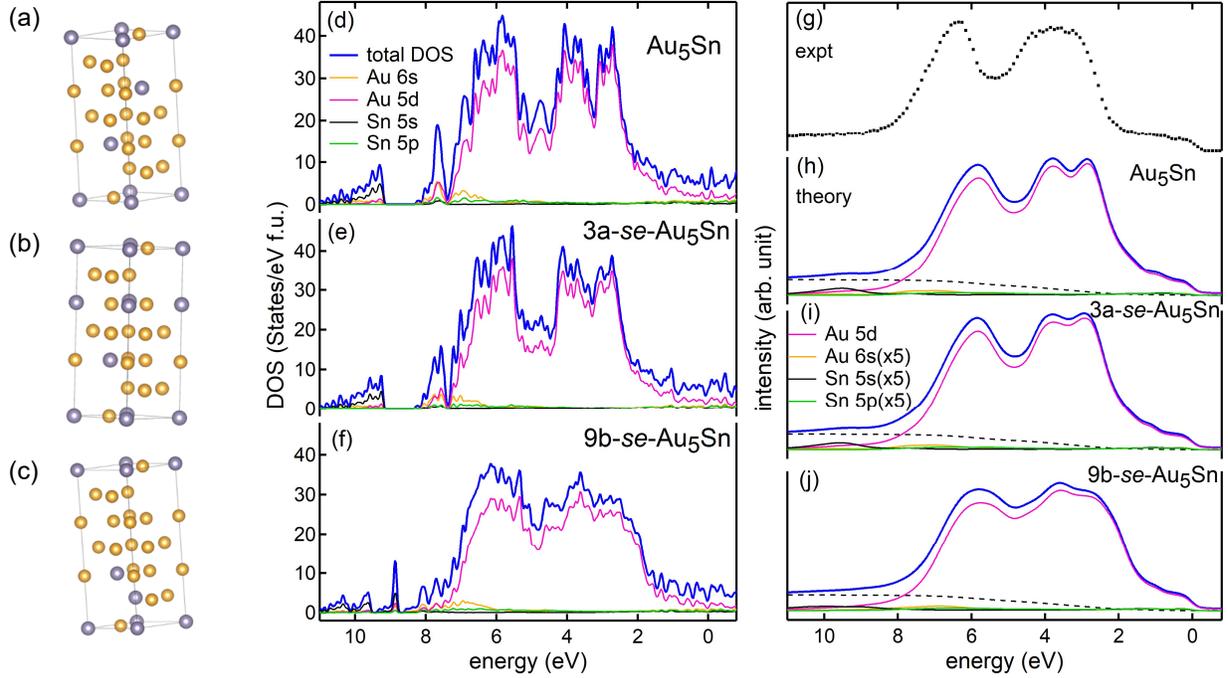} 
 	\vskip -25mm
 	\caption{ The crystal structure of (a) Au$_{5}$Sn, (b) 3a-se-Au$_5$Sn $i.e.$ site-exchange of 3a Au with Sn and (c) 9b-se-Au$_5$Sn $i.e.$ site-exchange of 9b Au with Sn. Au and Sn atoms are shown by yellow and grey filled circles, respectively. 
 			~ The calculated total and PDOS of  (d) Au$_{5}$Sn, (e) 3a-se-Au$_5$Sn and (f) 9b-se-Au$_5$Sn using density functional theory 
 			~and  comparison of the (g) experimental XPS VB of Au$_{5}$Sn with the calculated VB of (h) Au$_{5}$Sn,  (i) 3a-se-Au$_5$Sn and  (j) 9b-se-Au$_5$Sn. } 
 	\label{Au5Sn_DOS}
 \end{figure} 

\underline{{\bf XPS valence band of Au$_5$Sn: experiment and theory:}}  It is clear from the above discussion that the VB corresponding to a composition of Au$_{5.1}$Sn ($\approx$Au$_5$Sn)  for $t_{d}$= 16 min at $T_{A}$= 518~K (Table~\ref{heattreatment}) would not have any admixture from the substrate because of the increased thickness of the layer.  We choose the VB of this composition to compare with the theory of bulk Au$_5$Sn [whose structure is reported in literature\cite{Osada74}, see Fig.~\ref{Au5Sn_DOS}(a)] because the experimental composition is close to  Au$_5$Sn.  
~Fig.~\ref{Au5Sn_DOS}(d) shows that the total DOS of Au$_{5}$Sn 
~calculated  using this structure
~is dominated by Au 5$d$ states with small contribution of Au 6$s$ between 6 to 8~eV. In contrast, Sn 5$p$ has small intensity in the 0 to 8~eV region, whereas the peak at 9.5~eV  mainly arises from Sn 5$s$ states. 
  The  experimental VB of Au$_{5}$Sn in Fig.~\ref{Au5Sn_DOS}(g) shows good agreement with the calculated VB in  Fig.~\ref{Au5Sn_DOS}(h). The calculated VB  exhibits a dominant contribution from Au 5$d$ and Sn 5$s$ states in 0 to 8~eV and 8.5 to 10.8~eV ranges, respectively. The peak at 6.3 eV in the experimental VB is reproduced by a peak of similar shape at 5.8 eV. The broad peak  centered around 4 eV also has its counterpart in the theory. The agreement is thus  satisfactory, except for a dip that is clearly visible in theory at 3.3 eV, but is absent in experiment. 

 To find a possible reason for this disagreement, we have performed the calculations  with exchange of Au and Sn. The DOS and the PDOS for 3a-se-Au$_5$Sn [exchange of one Sn in 3a and one Au in 3a sites, Fig.~\ref{Au5Sn_DOS}(b)] and  9b-se-Au$_5$Sn [exchange of one Sn in 3a and one Au in 9b sites, Fig.~\ref{Au5Sn_DOS}(c)] are shown in Figs.~\ref{Au5Sn_DOS}(e,f). The dip is reduced in the former and is clearly absent in the latter. The reduction or near-absence of the dip is also evident in the calculated VB in Figs.~\ref{Au5Sn_DOS}(i) and Figs.~\ref{Au5Sn_DOS}(j), respectively. 
~Thus, anti-site defects could be a reason for the absence of the dip in the experimental VB. The broadening of the VB peaks due to disorder has been observed in other intermetallic compounds  earlier.\cite{Sadhukhan19}  In contrast to AuSn, the influence of  defect/disorder effect is more pronounced in Au$_5$Sn, possibly because it is obtained by high temperature treatment (and subsequent cooling to RT)  while AuSn is grown at RT.


     \subsection{ARPES and LEED study of the surfaces of Au-Sn  after heat treatment}

     \begin{figure*}[htb]
     	\vskip -20 mm
     	\centering
     	\includegraphics[width=165mm]{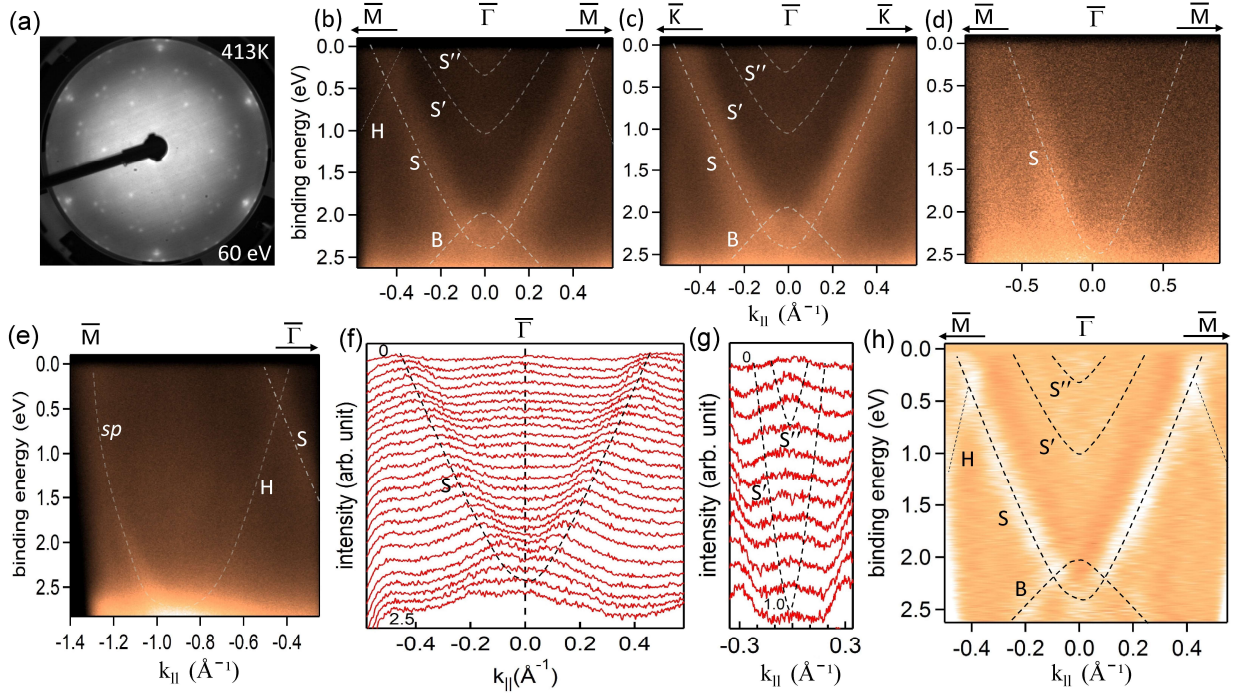} 
     	\vskip -15 mm
     	\caption{ (a) p(3\,$\times$\,3)R15$^{\circ}$ LEED pattern  for the Au$_{3.6}$Sn  surface  obtained by $t_{d}$= 16 min, $T_{S}$= 413~K.  ARPES spectra of this surface around $\overline{\Gamma}$ point (b) along $\overline{\Gamma}$-$\overline{\rm M}$ direction, (an arrow below a high symmetry point ($\overline{\rm M}$, $\overline{\rm K}$ or $\overline{\Gamma}$) indicates the direction along that high symmetry point)
		and (c) $\overline{\Gamma}$-$\overline{\rm K}$ direction of the substrate Brillouin zone.
     	~	(d) band dispersion around $\overline{\Gamma}$ along $\overline{\Gamma}$-$\overline{\rm M}$ direction at 40.8~eV (h$\nu$) (e) ARPES spectra around $\overline{\rm M}$ point ($k$= 1.26\AA{}$^{-1}$). (f) Momentum distribution curves of (b) drawn at 0.1 eV step, width 0.05 eV from 0 to 2.5~eV BE, (g) an expanded area of (f) from 0 to 1~eV, (h) the second derivative ARPES spectra of (b). All the dispersing features in (b-e) are drawn by white dashed and dotted lines for guide to eye.}
     	     	\label{arpes_ts413K}
     \end{figure*}
     
   The band structure of the Sn based surface alloys constitutes an interesting study.   Chuang {\it et al.} predicted that surface alloys can be used  to obtain 2D topological insulators.\cite{Chuang16} Our recent study of thin Au$_2$Sn layer prepared by small amount of Sn deposition equivalent to about a monolayer 
  	~ at $T_S$ $\approx$400 K showed a 
  	~linear band that meets at the Fermi level in the zone-center  for a complicated three domain surface phase.
  	\cite{Sadhukhan_archive} 
~ Here,  we report the results of our ARPES study for three different surface structures such as  
~$p$(3\,$\times$\,3)R15$^{\circ}$, ($\sqrt{3}$$\times$$\sqrt{3}$)R30$^{\circ}$ and a mixed phase (Table~\ref{heattreatment}).

The $p$(3\,$\times$\,3)R15$^{\circ}$ phase, whose LEED pattern is shown in Fig.~\ref{arpes_ts413K}(a), has been prepared by us using the following parameters $t_d$= 16 min and $T_S$= 413 K. This  was identified earlier as a Au-Sn alloy phase.\cite{Barthes81}  
~  The ARPES of this surface exhibits an electron like steeply dispersing parabolic band S around the $\overline{\Gamma}$ point along $\overline{\Gamma \rm M}$ direction [Fig.~\ref{arpes_ts413K}(b)] of the substrate Brillouin zone (BZ). The S band shows its minima at 2.5~eV and crosses the Fermi level (E$_{F}$) at 0.48\AA{}$^{-1}$. Moreover, this band overlaps with the modified $d$ band (denoted as B) below 2~eV BE. It is noteworthy to mention that the S band remains invariant along $\overline{\Gamma \rm K}$ direction [Fig.~\ref{arpes_ts413K}(c)] as well as for another photon  energy such as 40.8~eV [Fig.~\ref{arpes_ts413K}(d)]. This implies that this band is a surface state. Two small electron like bands S$^{\prime}$ and S$^{\prime\prime}$ are also identified at $\overline{\Gamma}$ along both $\overline{\Gamma \rm M}$ and $\overline{\Gamma \rm K}$ directions, although these are relatively weak. In addition to this, another band
~H has also been observed dispersing towards higher BE with increasing k$_{||}$ along $\overline{\Gamma \rm M}$ direction [marked by white dotted line in Fig.~\ref{arpes_ts413K}(b)] which bends into a parabolic band and is meeting with Au $sp$ band near $\overline{\rm M}$ point [Fig.~\ref{arpes_ts413K}(e)]. Fig.~\ref{arpes_ts413K}(f) presents the momentum distribution curves (MDCs) of Fig.~\ref{arpes_ts413K}(b) drawn in 0.1~eV step from E$_{F}$ to 2.5~eV BE taking average of MDCs within 0.05 eV width.
~An expanded part of Fig.~\ref{arpes_ts413K}(f) around $\overline{\Gamma}$ from 0 to 1 eV BE is shown in  Fig.~\ref{arpes_ts413K}(g), which provides an evidence about the presence of S$^{\prime}$ and S$^{\prime\prime}$ bands. Moreover, all the features identified in Fig.~\ref{arpes_ts413K}(b) are clearly visible in its second derivative image in Fig.~\ref{arpes_ts413K}(h) [shown by black dashed and dotted lines]. It may be noted that the absence of the linear bands around the $\overline{\Gamma}$ point for this surface indicates that these bands are specific to the (2113) phase.

\begin{figure}[tb]
	\vskip -20 mm
	\centering
	\includegraphics[width=190mm]{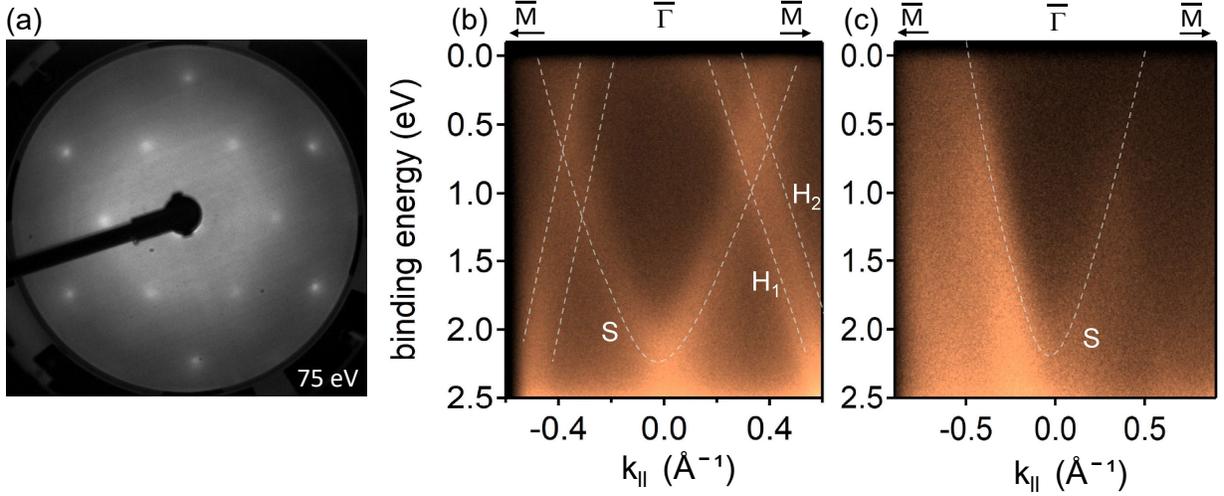} 
	\vskip -60mm
	\caption{ (a) The ($\sqrt{3}$$\times$$\sqrt{3}$)R30$^{\circ}$ LEED pattern of Au$_{5.1}$Sn using 75 eV primary beam energy  prepared with t$_{d}$= 16 min and  T$_{A}$= 518~K.  
		ARPES spectra of this phase along $\overline{\Gamma}$-$\overline{\rm M}$ direction  recorded with (b) 21.2 eV (c) 40.8 eV photon energies.} 
	\label{arpes_root3}
\end{figure}

 The well known ($\sqrt{3}$$\times$$\sqrt{3}$)R30$^{\circ}$ surface alloy phase is ubiquitous for the highest treatment temperatures. For low Sn deposition equivalent to about a monolayer, the composition of this phase is well known to be Au$_2$Sn.\cite{Maniraj18} On the other hand, for large Sn deposition here with $t_d$= 16-57 min, the  ($\sqrt{3}$$\times$$\sqrt{3}$)R30$^{\circ}$ phase is observed over a composition range above $R$= 5.  Fig.~\ref{arpes_root3}(b,c) shows the band structure of Au$_{5.1}$Sn layer with ($\sqrt{3}$$\times$$\sqrt{3}$)R30$^{\circ}$ surface symmetry [Fig.~\ref{arpes_root3}(a)] prepared with t$_{d}$= 16 min and  T$_{A}$= 518~K. 
~Two hole like bands H$_{1}$ and H$_{2}$, which  cross the Fermi level (E$_{F}$)  at $\approx$0.2 and 0.3~eV along $\overline{\Gamma \rm M}$ [Fig.~\ref{arpes_root3}(b)] direction of the substrate Brillouin zone. In addition to this, an electron-like sharp parabolic band (S) is observed around $\overline{\Gamma}$ point dispersing from 2.3~eV crossing the E$_{F}$ at ($\approx$) 0.45\AA{}$^{-1}$. The S band is also observed at different photon energy such as 40.8~eV [Fig.~\ref{arpes_root3}(c)]. This implies that the S band is surface related. The hole-like bands show close resemblance with the ($\sqrt{3}$$\times$$\sqrt{3}$)R30$^{\circ}$ phase of other systems such as  Sn/Ag(111).\cite{Osicki13} 
~ In  contrast, the electron-like band S is not observed.

\begin{figure}[htb]
	\vskip -5 mm
	\centering
	\includegraphics[width=120mm]{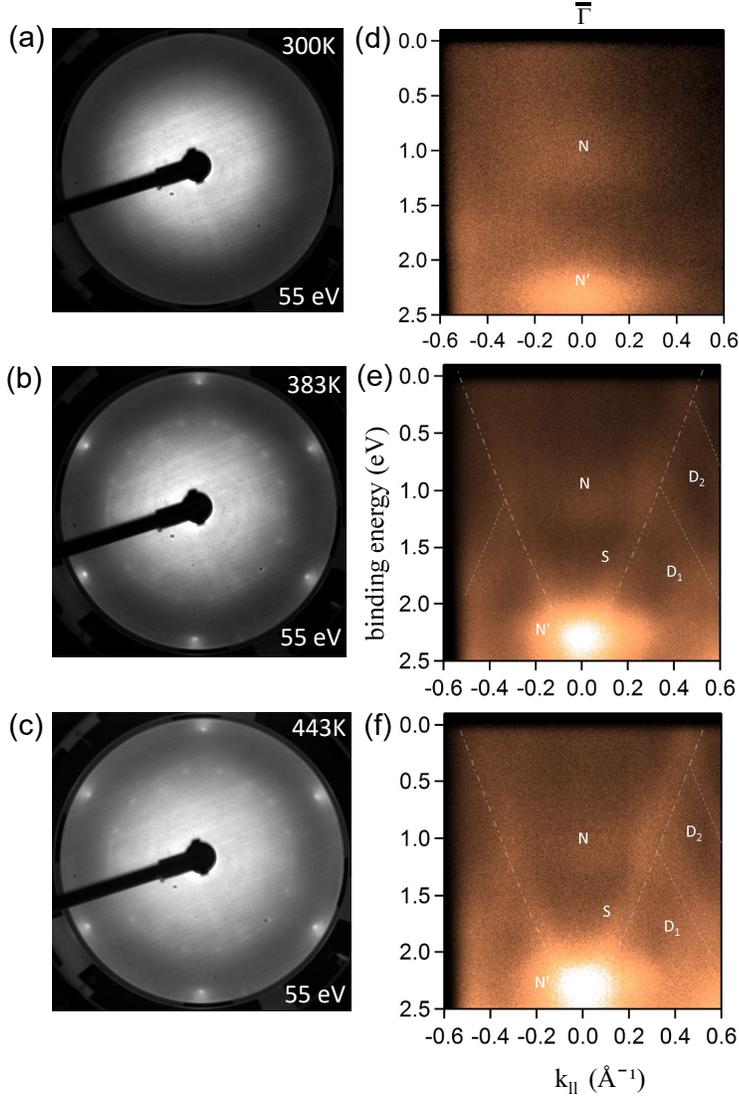} 
	\vskip -5 mm
	\caption{ LEED patterns of (a) 16 min Sn deposition on Au(111) at RT and post annealed at (b) $T_{A}$= 383~K, (c) $T_{A}$= 443~K 
		~at 55~eV electron energy. ARPES spectra of (d) 16 min Sn deposition on Au(111) at RT, and post annealed at (e) $T_{A}$= 383~K, and (f) $T_{A}$= 443~K 
		~around $\overline{\Gamma}$ point of substrate Brillouin zone along $\overline{\Gamma}$-$\overline{\rm M}$ direction. 
		~All the important features in the images are highlighted by white dashed lines as guide to the eye.} 
	\label{arpes_anneal1}
\end{figure}

Finally, it may be noted that in contrast to high temperature treatment,  the AuSn layer formed by  Sn deposition at RT  shows a diffuse LEED pattern with no sharp spots [Fig.~\ref{arpes_anneal1}(a)].  This indicates that the top surface of AuSn layer is not ordered, although the layer itself is ordered, as shown in Section IIIB. Since ARPES is more surface sensitive compared to XPS because of smaller $\lambda$$\approx$0.5 nm,~ dispersing bands are not observed; rather  two dispersionless features (N and N$^{\prime}$) are observed at  1.1~eV and 2.3 eV [Fig.~\ref{arpes_anneal1}(d)].  
~The Au-Sn layers obtained by post annealing in the range of  $T_{A}$= 383 K to 443 K has  composition ranging from Au$_{1.6}$Sn to Au$_{1.9}$Sn. These layers exhibit sharp LEED spots, but the pattern corresponds to a mixed phase [Fig.~\ref{arpes_anneal1}(b,c)]. LEED pattern corresponding to a mixed phase has also been reported for Sn/Cu(111)\cite{Liang11}. In this phase, a parabolic band around $\overline{\Gamma}$ along $\overline{\Gamma \rm M}$ direction of the   BZ is observed  [Fig.~\ref{arpes_anneal1}(e,f)], which is similar to the S band as presented in Fig.~\ref{arpes_ts413K}(b). In addition to this, some weak bands (D$_{1}$, D$_{2}$) that disperse to higher BE with increasing k$_{||}$ are also observed (white dotted lines). These bands become more prominent as $T_A$ is increased, as shown by Fig.~\ref{arpes_anneal1}(e,f).

 It may be noted that although LEED patterns provide information about the   unit cell symmetry,  the task of obtaining the atomic positions is complicated,  in particular  due to surface alloying. In case of p(3\,$\times$\,3)R15$^{\circ}$ that occurs for the composition range of Au$_{2.4}$Sn to Au$_{3.6}$Sn,  the bulk structure is not known since it does not exist in stable form. Moreover, sometimes, the surface need not be bulk terminated and might exhibit a reconstruction. Au(111) is a well known example that exhibits a complicated (22$\times$$\sqrt{3}$) reconstruction. Although the structure of the ($\sqrt{3}$$\times$$\sqrt{3}$)R30$^{\circ}$ phase with Au$_2$Sn composition is known\cite{Maniraj18}, the structure of the  surface where the layer composition is  $R$$\geq$5, albeit with same LEED symmetry, is not known.  For these reasons, we have  not been able to perform DFT calculations that require a structural model to understand  the origin of the ARPES bands discussed in this section.


\section{Conclusions}
 In this work, we have studied Au-Sn alloys formed by deposition of Sn on Au(111) up to thickness of the order of 10 nm under the following conditions:
	 (i) deposition at RT for different time ($t_d$); and (ii) deposition at elevated temperature ($T_S$)  and (iii) post annealing for different  $T_A$ at same $t_d$. In case of (i), Au 4$f$ and Sn 4$d$ core-level XPS show three different components, namely from the layer, interface and the substrate. The substrate and interface  components decrease while the layer component increases with $t_d$. 
	 ~For deposition at RT, the composition turns out to be AuSn, and the thickness is about 11~nm for $t_d$= 57 mins. The XPS valence band of the AuSn layer is in good agreement with theory, exhibits strong hybridization between Au 5$d$ and Sn 5$s$, 5$p$ states.  
	 ~ Both (ii) and (iii) lead to Au enrichment or Sn dilution due to inter-diffusion of Au and Sn. The AuSn related layer at RT decreases in intensity, whereas the  interface component increases in intensity with increasing heat treatment. Concomitantly, the Au-Sn layer width increases and this has been estimated by the decay of the substrate component in XPS. For (iii), the bulk structures such as AuSn and Au$_5$Sn are stable in a sizable temperature range and  transition from the former  to the latter  exhibits an activated behavior with an activation energy of about 9 kcal\,mol$^{-1}$. 
~Comparison of the theoretical results with the XPS VB  indicates that the influence of anti-site defects is much more pronounced in Au$_5$Sn compared to AuSn.
~Even though the nano-meters thick layer of AuSn grown at RT is  ordered, its surface   is disordered showing diffuse LEED pattern and no dispersing bands. Surface ordering is observed after heat treatment with appearance of p(3$\times$3)R15$^{\circ}$ surface phase that transforms to ($\sqrt{3}$$\times$$\sqrt{3}$)R30$^{\circ}$  phase at higher temperature. A nearly free electron-like parabolic surface state is observed for the p(3$\times$3)R15$^{\circ}$  phase, while  both electron-like and hole-like bands appear in the ($\sqrt{3}$$\times$$\sqrt{3}$)R30$^{\circ}$ phase. 

\section{Acknowledgments}
T. R. and A. C. thank D. Das, P. A. Naik and A. Banerjee for support and encouragement and the Computer Centre of   RRCAT, Indore for providing the computational facility. T. R. thanks Prof. M. Shirai for providing computational facility in Research Institute of Electrical Communication (RIEC), Tohoku University, Japan. 
~~\\

\noindent$^{*}$Present address: Research Institute of Electrical Communication (RIEC),Tohoku University, Sendai 980-8579, Japan.

\newpage
\setcounter{figure}{0}
\renewcommand{\figurename}{Fig.~S}
\setcounter{table}{0}
\renewcommand{\tablename}{Table.~S}

{\bf{\begin{center}Supplementary material for manuscript entitled\\ Electronic structure of Au-Sn compounds grown on Au(111)
\end{center}}}
\begin{center}{Pampa Sadhukhan$^1$, Sajal Barman$^1$, Tufan Roy$^{2,3,*}$, Vipin  Kumar Singh$^1$,  Shuvam Sarkar$^1$,  Aparna Chakrabarti$^{2,3}$ and Sudipta Roy Barman$^{1}$}
\end{center}

\affiliation{$^1$UGC-DAE Consortium for Scientific Research, Khandwa Road, Indore 452001, Madhya Pradesh, India}
\affiliation{$^2$Homi Bhabha National Institute, Training School Complex, Anushakti Nagar, Mumbai  400094, Maharashtra, India}
\affiliation{$^3$Theory and Simulations Laboratory,  Raja Ramanna Centre for Advanced Technology, Indore 452013, Madhya Pradesh, India}



\noindent{\bf This Supplementary material contains three figures and  five tables.} 

\begin{figure*}[htb]
	\centering
	\includegraphics[width=170mm]{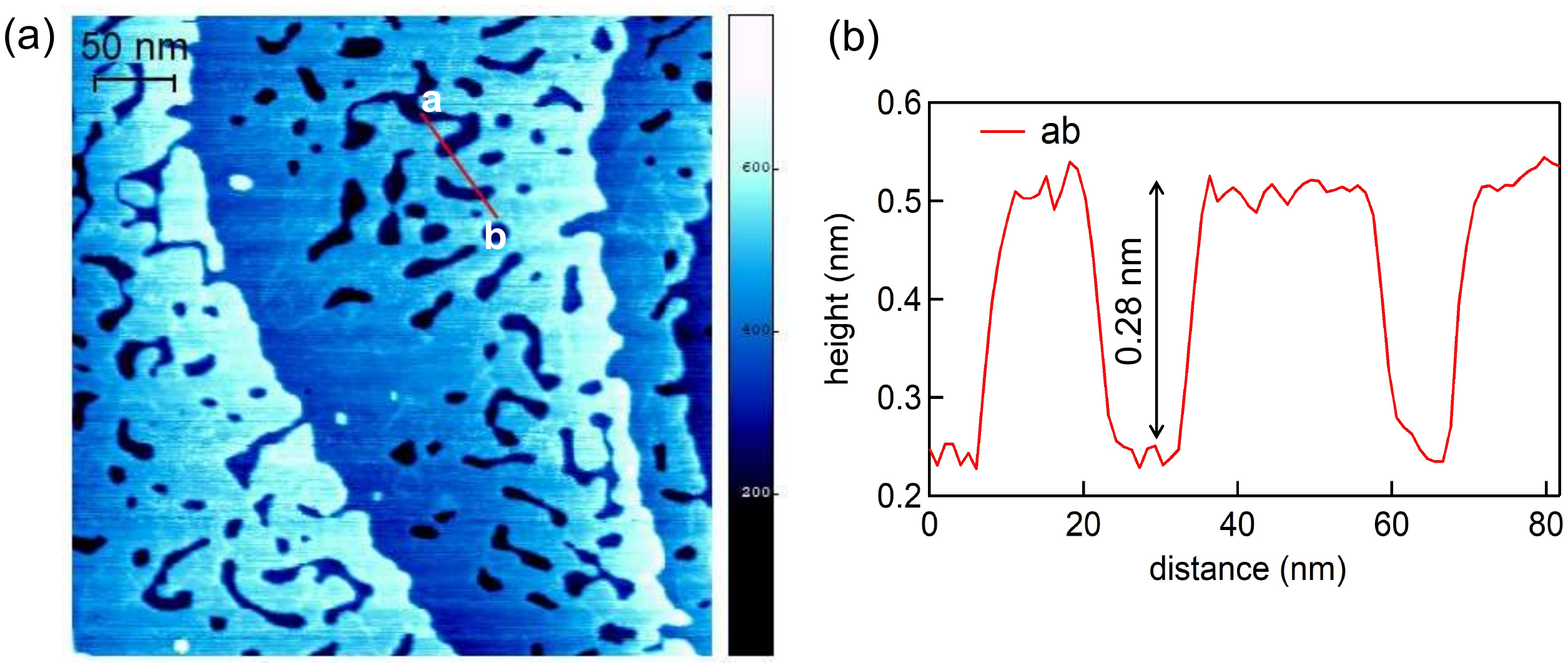} 
	\vskip -35mm
	\caption{(a) STM image of AuSn layer grown on Au(111) at RT, using I$_T$= 0.3 nA , U$_T$= 1.2 V, (b) height profile along ab of (a).} 
	\label{plasmon}
\end{figure*}

\begin{figure*}[htb]
	\centering
	\includegraphics[width=170mm]{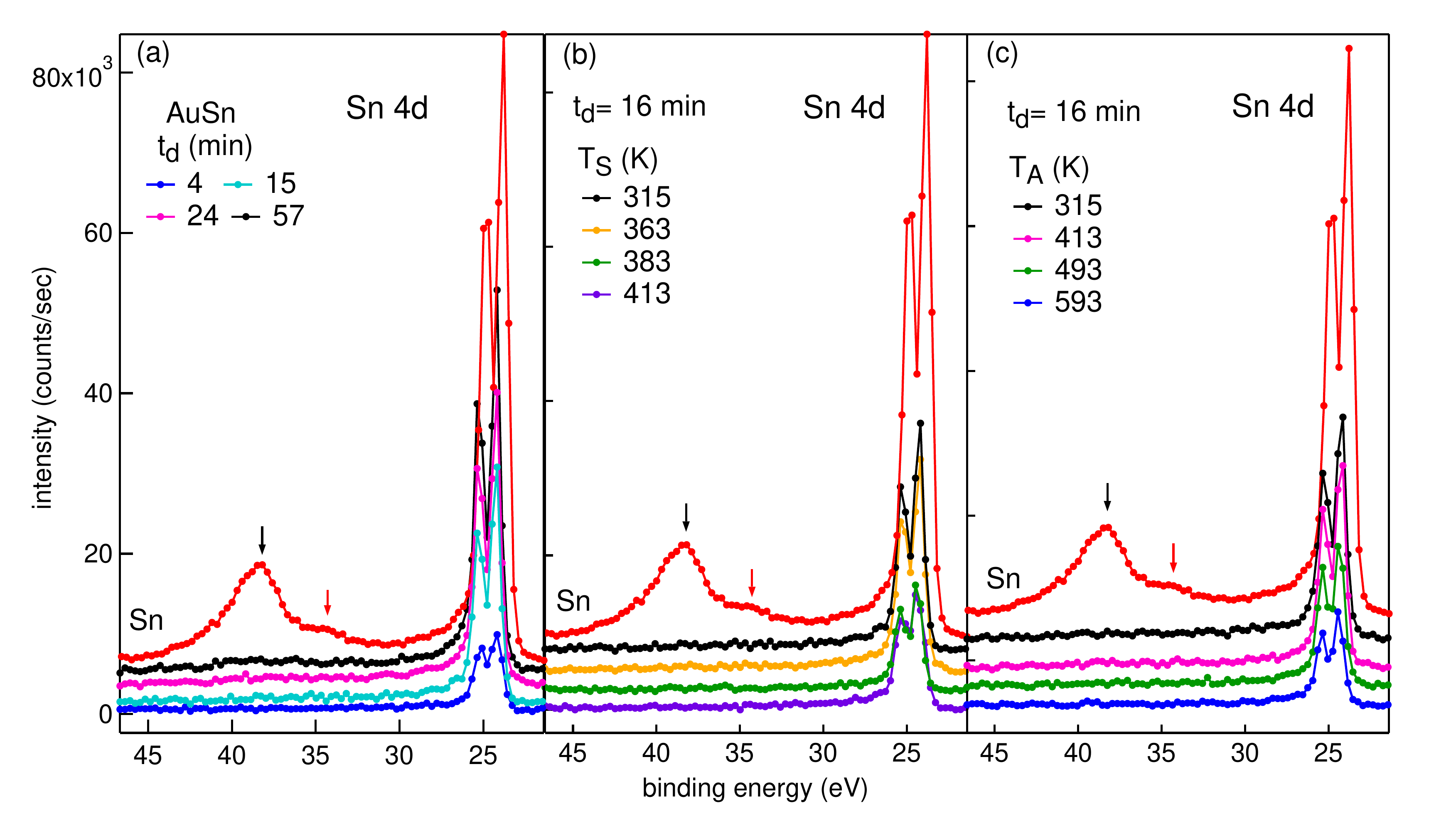}
	\caption{Sn 4$d$ XPS core-level spectrum showing the  first bulk plasmon (black arrow) and the surface plasmon (red arrow) compared with that of (a) AuSn for different $t_{d}$ at RT, Au-Sn layer (b) as a function of $T_S$ for $t_d$= 16 min and (c) as a function of $T_A$ for $t_d$= 16 min.} 
	\label{plasmon}
\end{figure*}

~~\\
~~\\
~~\\
~~\\
~~\\
~~\\

~~\\
~~\\
~~\\
\begin{figure*}[htb]
	\centering
	\includegraphics[width=130mm]{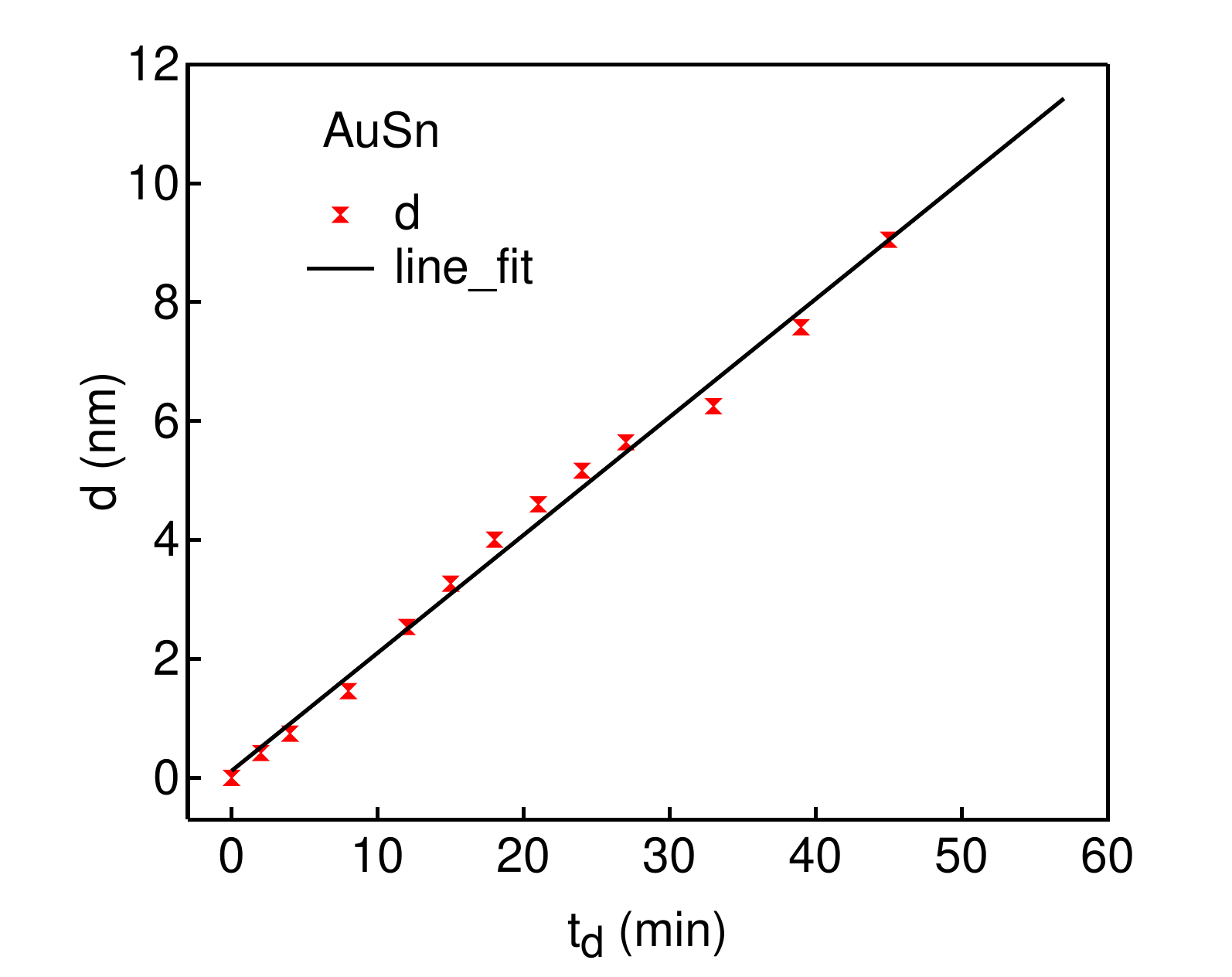} 
	\caption{Variation of AuSn layer thickness as a function of Sn deposition time (t$_d$) on Au(111) substrate at RT.} 
	\label{plasmon}
\end{figure*}

~~\\
~~\\
~~\\
~~\\
~~\\
~~\\

~~\\
~~\\
~~\\
~~\\
~~\\
~~\\
~~\\
~~\\
~~\\

~~\\
~~\\
~~\\
~~\\
~~\\
~~\\
~~\\
~~\\
~~\\

~~\\
~~\\
~~\\
\begin{table}  
	\caption{The  parameters obtained from least square fitting of Au 4$f$ core-level spectra for Sn deposition on Au(111) at RT such as binding energy ($E_B$), Gaussian FWHM ($\sigma$), Doniach-Sunjic asymmetry parameter ($\alpha$)\cite{Doniach70} of Au$_{\rm B}$, Au$_{\rm A1}$ and Au$_{\rm A2}$ components (see Section III.A. of the manuscript).}
	\centering
	\begin{tabular}{|c|c|c|c|c|c|c|c|c|c|c|c|c|c|c|c||c|} \hline 
		\multicolumn{1} {|c|}  {$t_d$ (min)  } & \multicolumn{3} {|c|}  {$E_B$ $\pm$ 0.05 (eV) }  & \multicolumn{3} {|c|}  {$\sigma$ (eV) }   & \multicolumn{1} {|c|} {$\alpha$ } & \multicolumn{1} {|c|} {Au:Sn} \\ \hline
		& Au$_{\rm B}$(4f$_{7/2}$) &  Au$_{\rm A2}$(4f$_{7/2}$)  & Au$_{\rm A1}$(4f$_{7/2}$) &  Au$_{\rm B}$ & Au$_{\rm A2}$ & Au$_{\rm A1}$ & & \\  
		0 & 84 & - &  - & 0.34 & - &-& 0.043  & \\
		2 & 84 & 84.39 & - & 0.34 & 0.34 & - & 0.048 & 1.13 \\
		4 & 84 & 84.39 & 84.76 &  0.34 & 0.34 & 0.34 & 0.04  & 1.42 \\  
		8 & 84 &  84.39  & 84.82 &  0.35 & 0.40 & 0.44 & 0.047  & 1.33 \\
		12 & 84 &  84.34  & 84.84 &  0.34 & 0.43 & 0.49 & 0.047  & 1.34 \\
		15 & 84 &  84.33  & 84.86 &  0.35 & 0.47 & 0.47 & 0.05  & 1.29 \\
		18 & 84 &  84.33  & 84.87 &  0.355 & 0.5 & 0.46 & 0.052  & 1.23 \\
		21 & 84 &  84.33  & 84.87 &  0.35 & 0.5 & 0.45 & 0.054  & 1.16 \\
		24 & 84 &  84.34  & 84.89 &  0.35 & 0.58 & 0.44 & 0.056  & 1.12 \\
		27 & 84 &  84.37  & 84.88 &  0.35 & 0.5 & 0.44 & 0.059  & 1.06 \\
		33 & 84 &  84.39  & 84.87 &  0.35 & 0.4 & 0.45 & 0.06  & 0.94 \\
		39 & 84 &  84.39  & 84.87 &  0.35 & 0.38 & 0.45 & 0.059  & 0.9 \\
		45 & 84 &  84.39  & 84.87 &  0.35 & 0.41 & 0.45 & 0.059  & 0.9 \\
		57 & - &  84.39  & 84.87 &  - & 0.49 & 0.46 & 0.059  & 0.9 \\ \hline

	\end{tabular}
	\label{fitting_parameters}
\end{table}

~~\\
~~\\
~~\\
~~\\
~~\\
~~\\

~~\\
~~\\
~~\\

\begin{table}  
	\caption{The  parameters  obtained from least square fitting of Sn 4$d$ core-level spectra for Sn deposition on Au(111) at RT for components such as Sn$_{\rm A1}$  and Sn$_{\rm A2}$ (see Section III.A. of the manuscript) as well as that of   Sn metal  (Sn$_{\rm B}$). } 
	\centering
	\begin{tabular}{|c|c|c|c|c|c|c|c|c|c|c|c|c|c|c|c|} \hline 
		\multicolumn{1} {|c|}  {$t_d$ (min)  } & \multicolumn{3} {|c|}  {$E_B$ $\pm$ 0.05 (eV) }   & \multicolumn{1} {|c|} {$\alpha$ } \\ \hline
		& Sn$_{\rm B}$(4d$_{5/2}$) &  Sn$_{\rm A2}$(4d$_{5/2}$)  & Sn$_{\rm A1}$(4d$_{5/2}$)  &  \\  
		Sn metal & 23.9 & - & -&  0.095  \\
		2 & - & 24.33 & 23.96  & 0.115  \\
		4 & - & 24.38 & 24  & 0.109  \\  
		8 & - & 24.32 & 24.18  & 0.109  \\
		12 & - &  24.3  & 24.13  & 0.109 \\
		15 & - &  24.37  & 24.16  & 0.103  \\
		18 & - &  24.43  & 24.18 & 0.102  \\
		21 & - &  24.49  & 24.18 & 0.101 \\
		24 & - &  24.5  & 24.18 & 0.099 \\
		27 & - &  24.5  & 24.18  & 0.101 \\
		33 & - &  -  & 24.18  & 0.145 \\
		39 & - &  -  & 24.18 & 0.142  \\
		45 & - &  -  & 24.18 & 0.143 \\
		57 & - &  -  & 24.18  & 0.142  \\ \hline

	\end{tabular}
	\label{fitting_parameters}
\end{table}

\begin{table}  
	\caption{The preparation parameters and all the Au-Sn compositions obtained by high temperature  ($T_S$) deposition for  $t_d$=16 mins.  The  Knudsen cell temperature (1140$\pm$5 K), the normal incidence geometry and distance of the sample with respect to the cell are unchanged.  Each entry in the $t_d$ column indicates a deposition on  sputter-annealed clean Au surface.}
	\begin{tabular}{|c|c|c|c|} \hline 
		\multicolumn{2} {|c|}  {Preparation parameters} &  \multicolumn{1} {|c|}  {Composition} \\  
		t$_d$ (min) &  T$_S$ (K)   & \\ \hline  
		16 & 315 &  Au$_{1.1}$Sn   \\
		16 & 333 &  Au$_{1.4}$Sn   \\
		16 & 343 &  Au$_{1.4}$Sn   \\ 
		16 & 363 &  Au$_{1.6}$Sn   \\
		16 & 383 &  Au$_{2.4}$Sn \\
		16 & 413 &   Au$_{3.6 }$Sn\\ 
		16 & 433 &  Au$_{3.8}$Sn   \\
		16 & 453 &  Au$_{4.2}$Sn   \\
		16 & 473 &  Au$_{5.5}$Sn   \\
		16 & 493 &  Au$_{4.8}$Sn  \\
		16 & 513 &  Au$_{6.0}$Sn \\
		16 & 523 &  Au$_{6.5}$Sn \\
		16 & 543 &  Au$_{7.4}$Sn \\
		16 & 573 &  Au$_{9.5}$Sn \\ 
		16 & 593 &  Au$_{9.3}$Sn \\
		16 & 603 &  Au$_{9.9}$Sn \\
		16 & 618 &  Au$_{12.3}$Sn \\
		\hline

	\end{tabular}
	\label{heattreatment_TS}
\end{table}

\begin{table}  
	\caption{The preparation parameters and the Au-Sn compositions obtained by post annealing  for 10 mins at each $T_A$ after Sn deposition for  $t_d$=16 mins and  57 mins at RT.  The  Knudsen cell temperature (1140$\pm$5 K), the normal incidence geometry and distance of the sample with respect to the cell are unchanged.  Each entry in the $t_d$ column indicates a deposition on  sputter-annealed clean Au surface.}
	\begin{tabular}{|c|c|c|c|} \hline 
		\multicolumn{2} {|c|}  {Preparation parameters} &  \multicolumn{1} {|c|}  {Composition} \\  
		t$_d$ (min) &  T$_A$ (K) & \\ \hline  
		16 & 315 &  Au$_{1.1}$Sn  \\
		&   333 & Au$_{1.5}$Sn  \\
		&   343 & Au$_{1.5}$Sn  \\
		&   363 & Au$_{1.4}$Sn  \\
		&   383 &  Au$_{1.6}$Sn\\
		&   413 & Au$_{1.7}$Sn\\
		&   443 &  Au$_{1.9}$Sn \\
		&   473 & Au$_{3.4}$Sn \\
		&   493 & Au$_{4}$Sn \\
		&   518 & Au$_{5.1}$Sn  \\
		&   530 & Au$_{5.9}$Sn \\ 	
		&   543 & Au$_{5.3}$Sn \\
		&   558 & Au$_{6.4}$Sn \\
		&   573 & Au$_{6.3}$Sn   \\ 	
		&   593 & Au$_{6.5}$Sn \\
		&   603 & Au$_{9.3}$Sn \\
		&   613 & Au$_{11.7}$Sn   \\ 
		&   618 & Au$_{11.3}$Sn   \\ 
		\hline
		57 &  315 & Au$_{0.9}$Sn  \\
		&   493 & Au$_{3.3}$Sn  \\
		&   593 &  Au$_{6.1}$Sn  \\ \hline

	\end{tabular}
	\label{heattreatment}
\end{table}

\begin{table}  
	\caption{The Au-Sn compositions obtained by the Sn deposition for different $t_d$ in the range of 10-20 mins at $T_S$= 493~K. Here, the composition (R) is essentially unchanged, the variation being $\pm$3\%, which is much less than the  error in $R$ (15\%, see Section II of main manuscript).}
	\begin{tabular}{|c|c|c|c|} \hline 
		\multicolumn{1} {|c|}  {Deposition time} &  \multicolumn{1} {|c|}  {Composition} \\  
		t$_d$ (min) & $R$\\ \hline  
		10 & Au$_{5.0}$Sn  \\
		13 & Au$_{5.2}$Sn  \\
		14 & Au$_{5.2}$Sn  \\
		15 & Au$_{5.2}$Sn  \\
		16 &  Au$_{4.8}$Sn\\
		18 & Au$_{4.9}$Sn\\
		20 &  Au$_{5.2}$Sn \\\hline

	\end{tabular}
	\label{heattreatment}
\end{table}

\end{document}